\documentclass[preprint,5p,times,twocolumn]{elsarticle}

\usepackage{amssymb}
\usepackage{amsmath}
\usepackage[T1]{fontenc}
\setcitestyle{super,open={},close={}}
\usepackage{changes}
\renewcommand{\added}[1]{{\color{black}#1}}
\renewcommand{\deleted}[1]{}

\begin{document}

\newcommand{\ones}{\mathbf{1}}
\newcommand{\T}{^\top}

\begin{frontmatter}


%%             \fntext[label3]{}

\newpageafter{abstract}

\title{PiNNwall: Heterogeneous Electrode Models from Integrating Machine Learning and Atomistic Simulation}

\author[inst1]{Thomas Dufils}

\affiliation[inst1]{organization={Department of Chemistry-\AA{}ngstr\"{o}m Laboratory, Uppsala University},%Department and Organization
            addressline={L\"{a}gerhyddsv\"{a}gen 1, BOX 538}, 
            city={Uppsala},
            postcode={75121}, 
            country={Sweden}}
            
\author[inst1]{Lisanne Knijff}

\author[inst1]{Yunqi Shao}

\author[inst1]{Chao Zhang}
\ead{chao.zhang@kemi.uu.se}

\begin{abstract}

 Electrochemical energy storage always involves the capacitive process. The prevailing electrode model used in the molecular simulation of polarizable electrode-electrolyte systems is the Siepmann-Sprik model developed for perfect metal electrodes. This model has been recently extended to study the metallicity in the electrode by including the Thomas-Fermi screening length. Nevertheless, a further extension to heterogeneous electrode models requires introducing chemical specificity which does not have any analytical recipes. Here, we address this challenge by integrating the atomistic machine learning code (PiNN) for generating the base charge and response kernel and the classical molecular dynamics code (MetalWalls) dedicated to the modelling of electrochemical systems, and this leads to the development of the PiNNwall interface. Apart from the cases of chemically doped graphene and graphene oxide electrodes as shown in this study, the PiNNwall interface also allows us to probe polarized oxide surfaces in which both the proton charge and the electronic charge can coexist. Therefore, this work opens the door for modelling heterogeneous and complex electrode materials often found in energy storage systems. 

\end{abstract}


%\item Heterogenous electrode modes are made simple with machine-learning response kernels.
%\item Surface roughness and chemical heterogeneity have opposite effects on Helmholtz capacitance.
%\item The PiNNwall interface developed here allows for studying graphene oxide with proton charge. 


\end{frontmatter}



\section{Introduction}

Electrochemical energy storage systems are indispensable components for building a sustainable and fossil-free society with infrastructures such as electric vehicles and energy grids. In particular, supercapacitors and batteries have attracted an ever-increasing attention in research going from materials chemistry to cell manufacturing. This is evinced by the 15,374 and 66,561 research articles published between 2020-2022 containing the keywords ``supercapacitors'' and ``batteries'' respectively (Source: the Web of Science), and highlighted by the 2019 Nobel Prize in Chemistry. On the other hand, to disentangle such complexity in these systems and to advance the field through 
fundamental insight, a physical approach is clearly needed. 

\added{As compared to battery systems, the capacitive charging process is the dominant one in supercapacitors.} Indeed, electric double-layer capacitors (EDLCs) store energy from the electrostatic adsorption of ions on the electrode surface, which leads to a rapid charge-discharge cycle~\cite{Conway_1991}. In this case, the charge-transfer rate is vanishingly small, and the electrode can be considered as an ideally polarizable electrode~\cite{Schmickler:2010th}. This means chemical reactions and chemisorptions
may be excluded from the setting~\cite{Le20}, therefore, force field-based classical molecular dynamics (MD) is sufficient to simulate EDLCs.

The standard model for describing the charge distribution of polarizable electrodes is the Siepmann-Sprik model\cite{siepmann_influence_1995}. It was improved by Reed and Madden \cite{reed_electrochemical_2007} to model perfect metal electrodes. Further improvements were done to account for the metallicity of the electrode material \cite{scalfi_semiclassical_2020}. This model has the advantage over other methods such as the image charge method~\cite{So21} to allow dealing with complex geometries, such as porous and disordered ones~\cite{2022.Jeanmairet}. 

\begin{figure}[ht]
\begin{center}
\includegraphics[width=\linewidth]{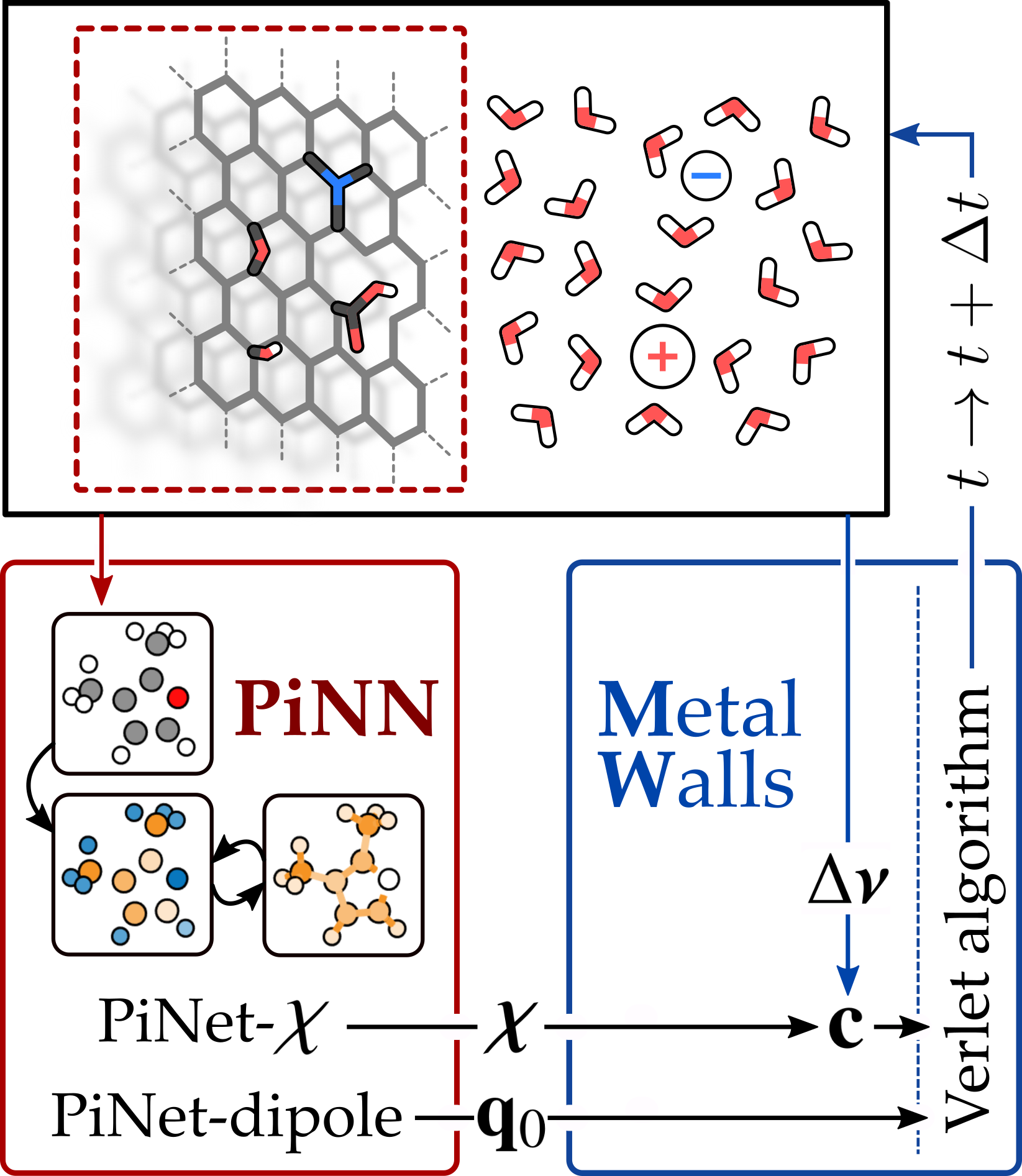}
\end{center}
\caption{\textbf{The flowchart of the PiNNwall interface}. The electrode structure is passed from MetalWalls to PiNN, which computes the charge response kernel $\boldsymbol{\chi}$ using PiNet-$\chi$, and the base charges of the electrode atoms $\mathbf{q}_0$ using PiNet-dipole. From the electrolyte configuration and the electrostatic boundary conditions, MetalWalls computes the potential on the electrode sites $\Delta\boldsymbol{\nu}$. By combining $\boldsymbol{\chi}$ and $\Delta\boldsymbol{\nu}$, MetalWalls generates the response charges $\mathbf{c}$ at electrode sites, computes forces and propagates the dynamics of the system using, for example, the Verlet algorithm. \label{flowchart_pinnwall}}
\end{figure}

 Despite being successful for describing both the perfect metal (PM) electrode and the Thomas-Fermi (TF) electrode, the Siepmann-Sprik model does not naturally account for chemical heterogeneity~\cite{Zhan17, 10.1002/eem2.12124, 2020.Bi, 2021.Pereira, Sato22, 2022.Bi}. This is also true when it comes to the local effects of electrode geometry and atomic lattice disorder on metallicity. To account for the impact of the chemical heterogeneity of the electrode material on the response charge distribution, our approach here is to integrate machine learning (ML) and atomistic simulation  with the PiNNwall interface, as shown in the Fig.~\ref{flowchart_pinnwall}. The purpose of this interface is to read the electrode structure from the classical MD code MetalWalls~\cite{marin2020metalwalls, Coretti22}, to compute the charge response kernel and the base charge with the atomistic ML code PiNN~\cite{2020_ShaoHellstroemEtAl} and then pass these info back to the MetalWalls for computing the response charge at electrode sites and propagating molecular dynamics simulations. By doing so, we can take advantage of both the efficient implementation of ML models in PiNN and the optimized computation of electrostatic interactions in MetalWalls. 

In the following, we will first outline the computational methods used in this study including the theoretical formulation. This is followed by the implementation and the validation of the PiNNwall interface to make sure of its technical soundness. Then, the PiNNwall interface is applied to several cases of chemically doped graphene and graphene oxide where the chemical heterogeneity becomes important. In particular, we have showcased an example of graphene oxide terminated with deprotonated carboxylic groups where both the electronic charge and the proton charge are present. Finally, we close up with a discussion of future works.

\section{Computational Methods}

\subsection{The Siepmann-Sprik model for polarizable electrode}
The basis of the Siepmann-Sprik model is to allow the electrode charges to fluctuate in response to the external potential. Each response charge of the electrode atoms follows a Gaussian distribution of magnitude $c_i$ centered on the position of the electrode atom $\textbf{R}_i$
\begin{equation}
    \rho_i(\mathbf{r})=c_i\left(\frac{\zeta_i}{\pi}\right)^{3/2}\exp^{-\zeta_i(\mathbf{r}-\mathbf{R}_i)^2}
\end{equation}
where $\zeta_i$ is an adjustable parameter related to the Gaussian width. 

The original model can be written as follows
\begin{equation}
    U = U_0 + U_{q_0-\Delta\nu} + \frac{1}{2} \mathbf{c}\T 
    \boldsymbol{\eta} \mathbf{c} +  \Delta \boldsymbol\nu\T \mathbf{c} \label{tot_energy}
\end{equation}
where $U_0$ corresponds to the energy of electrode atoms in absence of external potential (field). The term $U_{q_0-\Delta\nu}$ corresponds to the electrode-electrolyte interaction (so electrostatic interactions between the atomic charges of electrolyte atoms and the base charges $\mathbf{q}_0$ of electrode atoms plus their van der Waals interactions). $\boldsymbol\eta$ is the hardness kernel, describing the interaction between response charges and $\Delta \boldsymbol\nu$ is the potential generated by the electrolyte at the electrode atom sites. \added{It is worth noting that the formulation of the Siepmann-Sprik model shown here follows the linear response theory used in the chemical potential equalization method from York and Yang~\cite{York:1996ia}. This is different from other similar schemes~\cite{Sato19,2022.Bi}, in which the atomic electronegativtiy were introduced to determine the base charge distribution $q_0$. For historical developments on this topic and the subtle (yet important) difference in various schemes, we refer interested readers to our previous work~\cite{2022.Shao} and the atom-condensed Kohn-Sham DFT approximated to second order (ACKS2) paper~\cite{2013_VerstraelenAyersEtAl} for extensive discussions and references.}

This energy is minimized with respective to the response charge $\mathbf{c}$ at each MD time-step under the constraint of charge neutrality, which results in a linear relation between the response charge and the external potential as
\begin{equation}
    \mathbf{c} = \boldsymbol{\chi} \Delta \boldsymbol{\nu} \label{c_def}
\end{equation}
where $\boldsymbol\chi$ is the charge response kernel (CRK). It is related to the hardness kernel through~\cite{BR_88}
\begin{equation}
    \boldsymbol{\chi} = -\boldsymbol{\eta}^{-1} + \frac{\boldsymbol{\eta}^{-1}\ones\otimes\ones\T\boldsymbol{\eta}^{-1}}{\ones\T\boldsymbol{\eta}^{-1}\ones} \label{chi_def}
\end{equation}
where the second term of the right-hand side comes out from the charge neutrality constraint. 

The finite-field extension in the case of a constant external field $\mathbf{E}_0$ is straightforward, which leads to the solution of the response charge as
\begin{equation}
    \mathbf{c} = \boldsymbol{\chi} (\Delta \boldsymbol{\nu} - \mathbf{R}\mathbf{E}_0)
\end{equation}
It is worth noting that the external field $\mathbf{E}_0$ equals to the Maxwell field $\mathbf{E}$ under periodic boundary conditions (PBCs)~\cite{Zhang:2016cl}. 

\subsection{Response charge predictions from PiNet-$\chi$}

PiNet-$\chi$~\cite{2022.Shao} is a ML model based on PiNet for predicting the linear response function CRK by regressing the molecular polarizability, as implemented in PiNN code~\cite{2020_ShaoHellstroemEtAl}.

In this study, we used PiNet-$\chi$ which has been trained on the QM7b dataset~\cite{Yang19} to reproduce molecular polarizabilities computed from the density-functional theory (DFT)~\cite{Becke14} with B3LYP functional~\cite{becke1988density, lee1988development}. Thus it is suited to model electrode materials composed of the following elements: C, N, O, H, S, Cl, which will be sufficient to study graphene (or graphite) and its derivatives, being amorphous graphene, nitrogen-doped graphene or graphene oxides.

There are four different types of models provided by PiNet-$\chi$, namely the electronegativity equalization method (EEM)-type~\cite{Mortier85, Shankar:1986kb}, the Local-type~\cite{2022.Shao}, the EtaInv-type~\cite{2022.Shao} and the ACKS2-type~\cite{2013_VerstraelenAyersEtAl}. In the following, the essence of each model is summarized and more details can be found in Ref.~\citenum{2022.Shao}.

In the EEM-type model, the hardness matrix $\boldsymbol{\eta}$ is approximated by $\boldsymbol{\eta}_{e}$.  $\boldsymbol{\eta}_{e}$ contains environment-dependent on-site hardness parameters, as well as the Coulomb kernel due to electrostatic interactions. From this, $\boldsymbol{\chi}$ can be computed according to Eq.~\ref{chi_def}. 

In the Local-type model, the polarizability tensor is constructed as the sum of atomic contributions $\boldsymbol{\alpha}_{i}$. Then, the atomic contributions $\boldsymbol{\alpha}_{i}$ are constructed from atom-centered predictions $\boldsymbol{\chi}_{i}$ in a way that ensures translational and permutational invariance and rotational covariance. $\boldsymbol{\chi}_{i}$ can be seen as atomic contributions to the CRK, and are used to construct $\boldsymbol{\chi}$ in the end.

In the EtaInv-type model, $\boldsymbol{\chi}$ is constructed by predicting directly the softness matrix $\boldsymbol{\eta}^{-1}$. Besides the nearsightedness character of $\boldsymbol{\eta}^{-1}$, this type of models are computational efficient, since the need for a matrix inversion operation is bypassed.

Finally, in the ACKS2-type model, two quantities are predicted instead, namely $\boldsymbol{\chi}_{s}$ and $\boldsymbol{\eta}_{e}$. Here, $\boldsymbol{\chi}_{s}$ is constructed as a matrix that is local and trainable using symmetrized pairwise interactions. $\boldsymbol{\eta}_{e}$ is done in the same way as in the EEM model. These two predicted quantities can then be combined to construct $\boldsymbol{\chi}$ through the Dyson's equation, as shown in Ref.~\citenum{2022.Shao}.
\begin{equation}
    \boldsymbol{\chi} = \boldsymbol{\chi}_{s} \left [ \mathbf{I} - \boldsymbol{\eta}_{e} \boldsymbol{\chi}_{s} \right ] ^{-1}
\end{equation}

\subsection{Base charge predictions from PiNet-dipole}

PiNet-dipole~\cite{knijff2021machine} is a ML model based on PiNet as implemented in PiNN code~\cite{2020_ShaoHellstroemEtAl}. The principle behind the PiNet-dipole model is to regress dipole moment/polarization data instead of atomic charge data, as the latter can not be uniquely determined. 

Here, a variant of PiNet-dipole trained on the QM7b dataset~\cite{Yang19} was used, to be compatible with PiNet-$\chi$. The model was trained using the following loss function
\begin{equation}
    \mathcal{L} = \sum ^{n} _{i} || \mathbf{R}_{i}\mathbf{q}_{i} - \mathbf{M}_{i} || ^{2}_{2}
\end{equation}
where $\mathbf{R}_i$ is a $3\times N_i$ matrix of the atomic coordinates of the configuration $i$ for a molecular configuration containing $N_i$ atoms. $\mathbf{q}_i$ represents a column vector of the atomic charge, and $\mathbf{M}_i$ is the corresponding dipole moment.

During the charge prediction phase, the base charge $\mathbf{q}_0$ is obtained by
\begin{equation}
    \mathbf{q}_0 =  \mathbf{q} - \frac{\mathbf{1}\otimes \mathbf{1}\T\mathbf{q}}{\mathbf{1}\T\mathbf{1}}
\end{equation}.
This means that the total charge after charge prediction is evenly spread over all the atoms in the system, resulting in a zero total charge in $\mathbf{q}_0$.

In the case of protonated and deprotonated carboxyl groups, the total charge of $\mathbf{q}_0$ of each carboxyl group is either $+1$ or $-1$. This constraint was implemented by adjusting the base charge of carbon atom in the carboxyl groups. 

Details of the validation and the implementation of base charges predicted from PiNet-dipole can be found in the Section B of Supporting Information. 

\subsection{Molecular dynamics simulations with MetalWalls}
The MetalWalls code\cite{marin2020metalwalls, Coretti22} was used as the MD engine, which was built for simulating electrochemical systems with Siepmann-Sprik-type models. The box lengths in the different directions are L$_{x}$= 31.974 Å, L$_{y}$= 34.080 Å and L$_{z}$= 70.124 Å. We use 3D PBCs, with Ewald summation used to compute electrostatic interactions with a real-space cutoff of 15.99 Å, the same cutoff being used for the Lennard-Jones interactions.

The electrode consists in 7 graphene layers with an interlayer spacing of 3.354 Å, resulting in 2912 carbon atoms, which leaves a 50 Å space along the z direction for the electrolyte. For each dopant type, we investigated, on top of the pristine case, two surface coverages: 10\% and 20\%. Only the graphene layers at the interface with the electrolyte are functionalized. In the case of nitrogen substitution, the atoms are placed randomly under the constraint that two nearest neighbour atoms cannot be substituted. For the doping with epoxy and hydroxyl groups, we used the rules for the amorphous graphene oxide model described in Ref.~\citenum{liu2012amorphous}. Lennard-Jones parameters of electrode atoms were taken from the OPLS-AA force field \cite{jorgensen1996development} with the use of the Lorentz-Berthelot mixing rules to compute the cross pair parameters with the electrolyte.

The simulation setup for the case of graphene oxide with the carboxyl termination is very similar to the case of protonic double layer at metal-oxide/electrolyte interfaces, as studied previously with finite-field DFTMD~\cite{zhang2019coupling,jia2021origin}, in which two sides of electrode take the same amount but opposite types of proton charge. 

As for electrolyte, we used an aqueous potassium chloride solution with a concentration of 1 mol/L, whose initial configuration has been generated with fftool\cite{2021_AgilioPaduaGolovizninaEtAl} and PACKMOL \cite{martinez2009packmol}. This results in 1901 water molecules and 35 ion pairs. Water was modelled with the TIP3P model \cite{jorgensen1983comparison} and the ion models of aqueous K$^+$ and Cl$^-$ were taken from Ref.~\citenum{Joung2008}, which have been validated for high salt concentrations\cite{Zhang:2010zh}.  

The potential-dependence is controlled through the finite-field methods adapted to the Siepmann-Sprik model \cite{dufils2019simulating}, using $E$ field values corresponding to potential differences across the simulation cell of 0 and 2V.
Each simulation consists in an equilibration run of 2 ns followed by a production run of 10 ns. We used a timestep of 2 fs in the NVT (constant number of particles, constant volume, and constant temperature) ensemble using the Nos\'{e}-Hoover thermostat~\cite{1984.Nose, Hoover85} with a relaxation time of 0.1 ps and a temperature of 300 K.

\section{Implementation and validations of PiNNwall}

\subsection{Passing the charge response kernel from PiNN to MetalWalls}

\begin{figure}[ht]
\begin{center}
\includegraphics[width=0.8\linewidth]{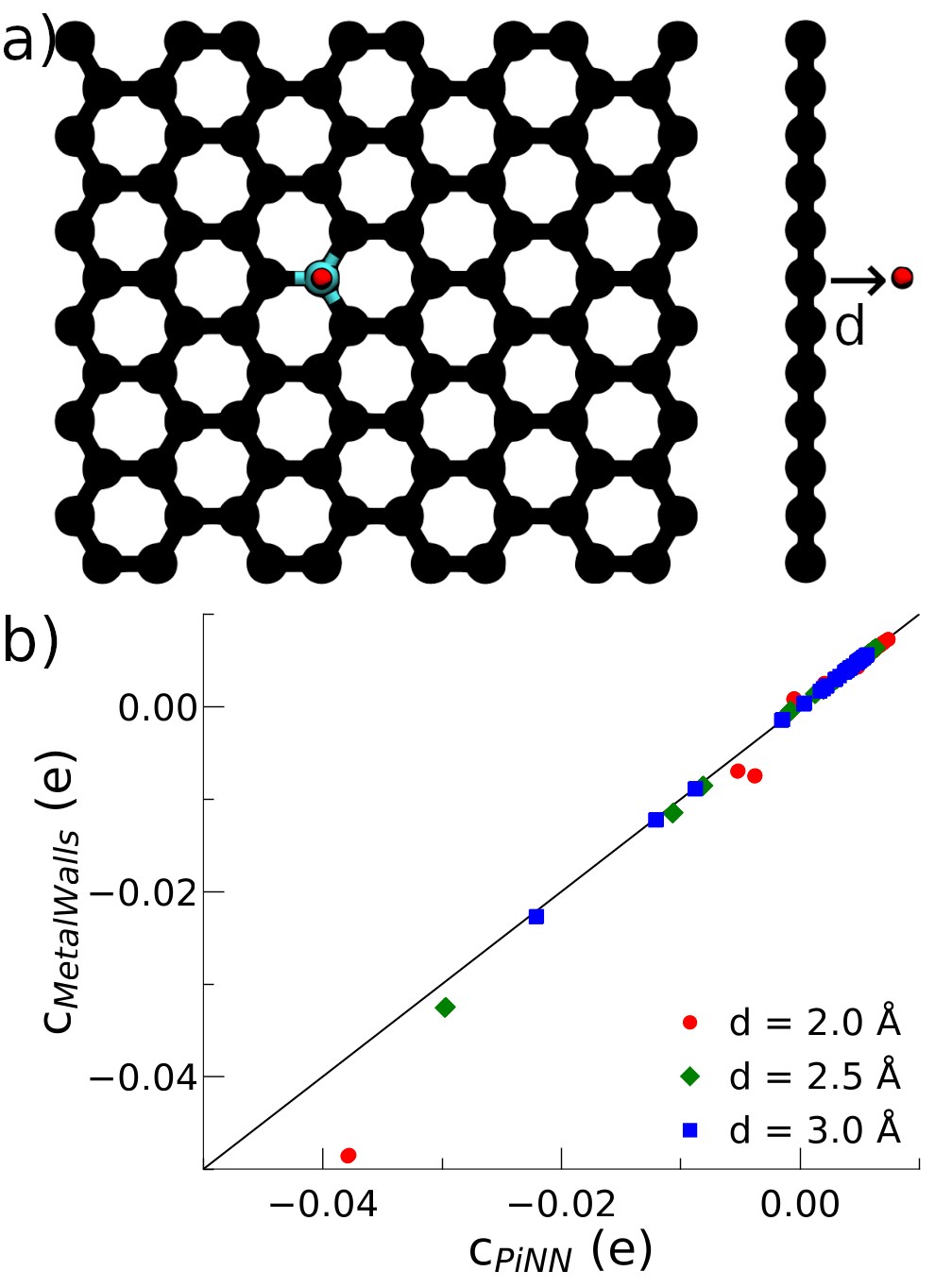}
\end{center}
\caption{\textbf{Passing the charge response kernel} a) A nitrogen-doped graphene layer with 3D PBCs. A unit test charge is put at a distance $d$ away from the surface on top of the defect. b) Response charges predicted by MetalWalls via the PiNNwall interface against the prediction from PiNN using the same kernel PiNet-$\chi$ (EEM).  \label{implementation}}
\end{figure}

 To test that the CRK $\boldsymbol\chi$ is properly passed to MetalWalls through the PiNNwall interface, we consider the system described on Fig.~\ref{implementation}a: a nitrogen-doped graphene layer with 3D PBCs. A unit test charge is placed away from the surface on top of the defect with a distance $d$. Then, the response charges were computed with the same EEM-type models using PiNet-$\chi$ and MetalWalls. Results are shown in Fig.~\ref{implementation}b. 
 One can see that the response charges agree very well with each other when the test charge is further away from the surface and  only atoms that are second neighbour to the defect and beyond are considered. This indicates that the CRK is indeed successfully passed from PiNet-$\chi$ to MetalWalls via the PiNNwall interface. The discrepancies in other cases actually come from how the Ewald summation for computing the electrostatic potential due to the test charge was implemented. In PiNN, the electrode-test charge interaction was computed as a point charge-point charge interaction; in MetalWalls, the electrode-test charge interaction was computed as a Gaussian charge-point charge interaction instead. Nevertheless, such difference is \added{immaterial}, and does not affect the passing of the CRK from PiNN to MetalWalls at all. \added{Indeed, one can obtain a perfect agreement when choosing a smaller Gaussian width (Section A in the Supporting Information). It is worth noting that there is no need to choose the Gaussian widths when using the PiNNwall interface for practical applications (Section 4) as the Gaussian widths that were optimized in PiNet-$\chi$ (EEM) will be passed to MetalWalls for computing the electrostatic interactions. Therefore, there is no risk of double-counting of the screening effect and the implementation is self-consistent.}

\subsection{Forces and the total energy from the charge response kernel}
In contrast to the original Siepmann-Sprik model and its TF variant, the CRK instead of the hardness kernel $\boldsymbol{\eta}$ is the key quantity used in PiNet-$\chi$. This means forces and the total energy in MetalWalls, that are formulated based on the hardness kernel, may not coincide with the CRK passed from PiNet-$\chi$. Thus we have to check the dependence on the hardness kernel of the quantities needed to run the MD and correct them if necessary. 

\begin{figure}[ht]
\begin{center}
\includegraphics[width=0.8\linewidth]{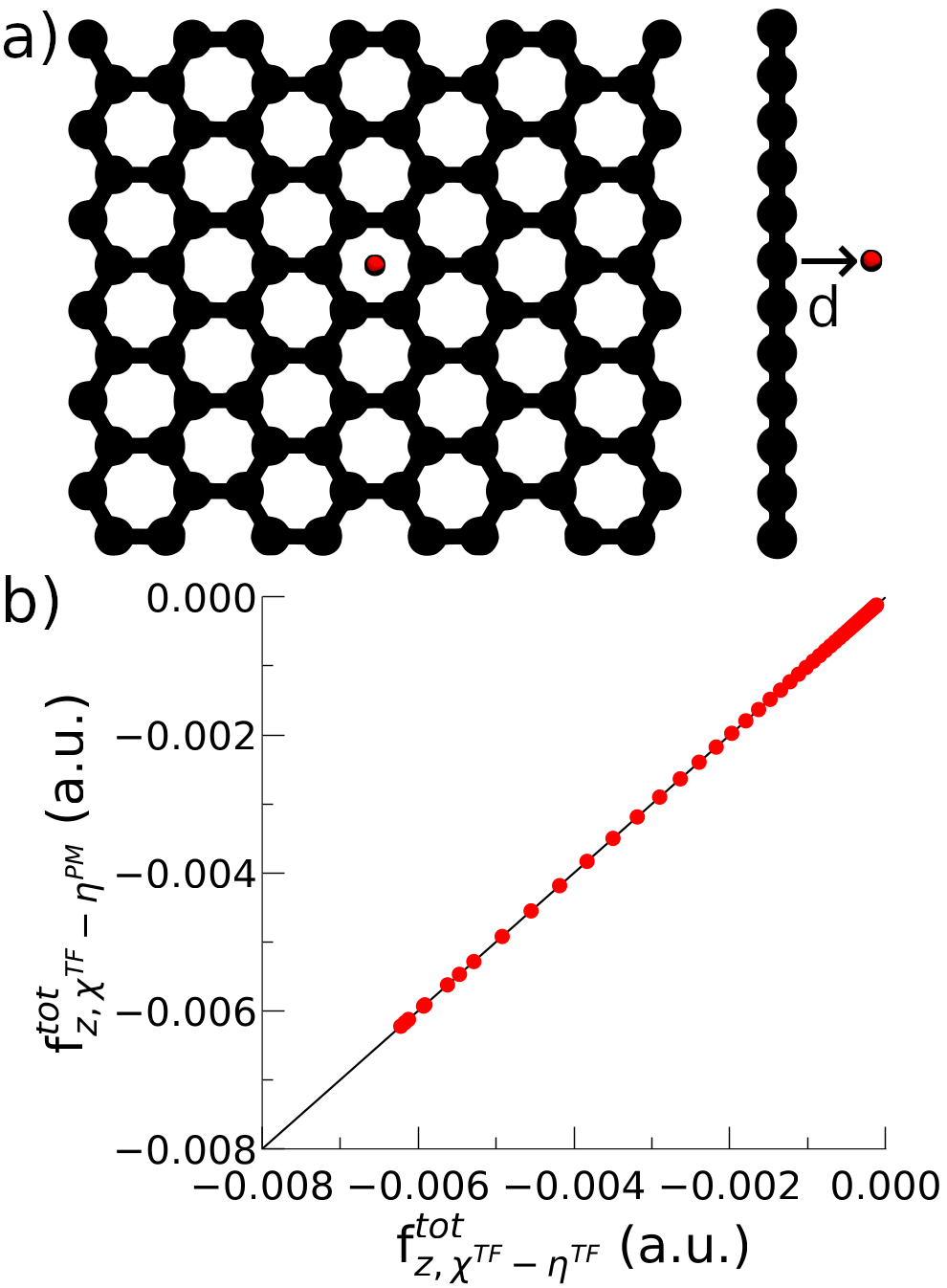}
\end{center}
\caption{\textbf{Forces from the charge response kernel}. a) A unit test charge is put on top of a graphene layer over the center of a six member ring, at a distance $d$ of the layer, ranging from 0.5 to 5.5 \AA.  b) Total contribution to the force acting on the test charge along the direction perpendicular to the surface. The subscript $\chi^{TF}-\eta^{TF}$ indicates both $\chi$ and $\eta$ come from the Thomas-Fermi model. The subscript $\chi^{TF}-\eta^{PM}$ indicates $\chi$ comes from the Thomas-Fermi model while $\eta$ results from the perfect metal electrode.
\label{forces_vs_eta}}
\end{figure}

To show whether these quantities depend on the hardness kernel or not, we use parameter sets of both PM and TF metals for constructing the hardness kernels $\boldsymbol\eta$ but only the parameter set of a TF metal for constructing the charge response kernel $\boldsymbol\chi$. Therefore, if the quantity in interest does not depend on $\boldsymbol\eta$, then the results will lie perfectly along the diagonal line in the parity plot. As all these tests were done with MetalWalls, we have used a system shown in Fig.  \ref{forces_vs_eta}a: a unit test charge is put on top of a graphene layer over the center of a six member ring, at a distance $d$ of the layer.

The forces caused by the interactions between the response charges and the electrolyte atoms at position $\mathbf{r}_i$ are given by
\begin{equation}
    \mathbf{f}_i = - \mathbf{c}^{\top} \nabla_{\mathbf{r}_i} \Delta \boldsymbol{\nu}  \label{forces_def}
\end{equation}

 According to Eq.~\ref{c_def}, the response charges depend only on the CRK. Since the external potential $\Delta \boldsymbol{\nu}$ does not depend on the hardness kernel either, neither should the forces. Indeed, as shown in Fig.~\ref{forces_vs_eta}b with the forces (acting along the perpendicular direction) are the same regardless what $\boldsymbol{\eta}$ is used.

\begin{figure}[ht]
\begin{center}
\includegraphics[width=0.7\linewidth]{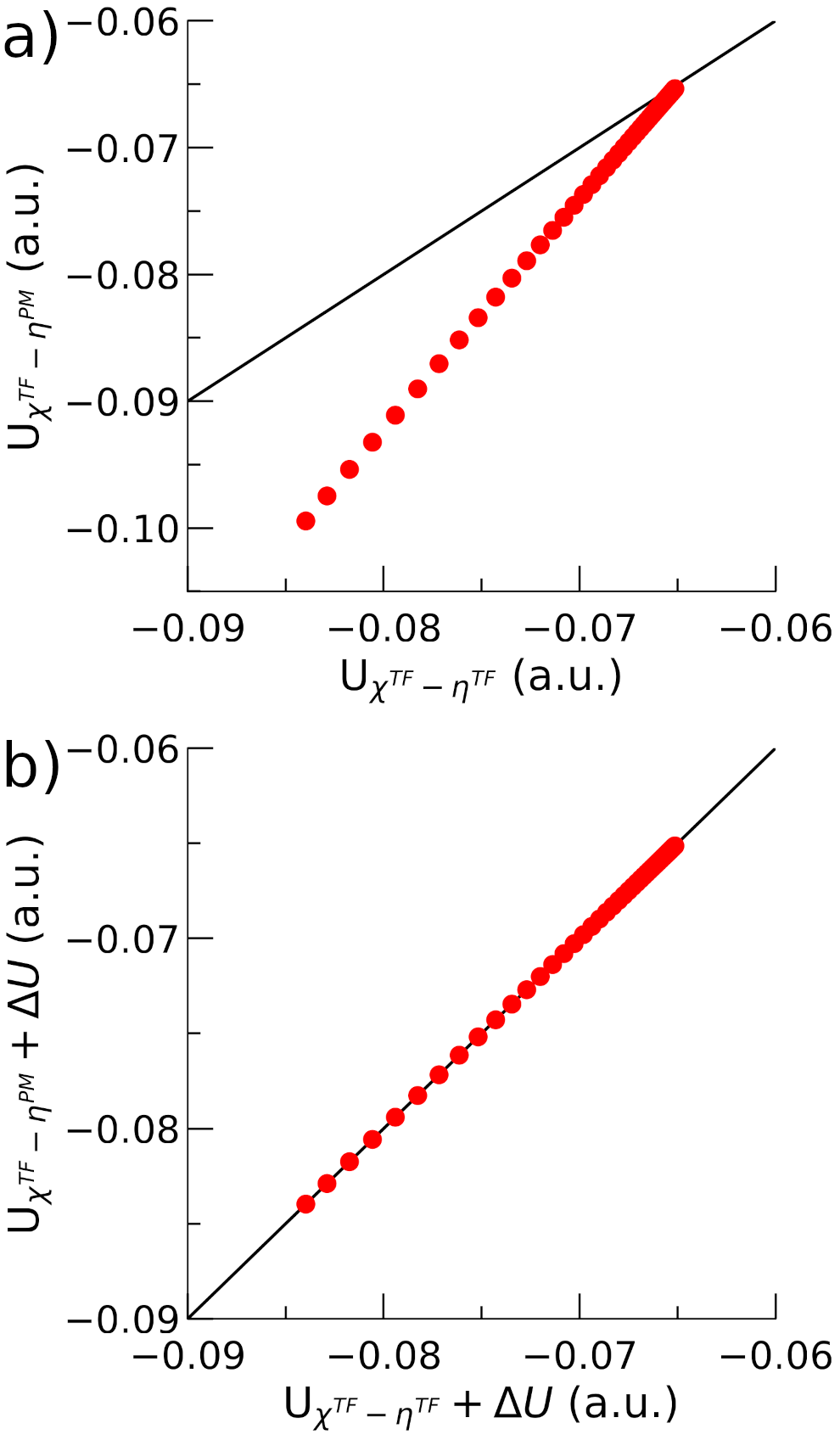}
\end{center}
\caption{\textbf{Hardness dependence of the total energy}. Using the simulation setup of figure \ref{forces_vs_eta}a).  a) Without the correction term in Eq.~\ref{correction_def}, the total energy expression depends on both the hardness and the charge response kernel. b) With the correction term in Eq.~\ref{correction_def}, the total energy depends only on the charge response kernel. The subscripts $\chi^{TF}-\eta^{TF}$ and $\chi^{TF}-\eta^{PM}$ follows the same convention used in Fig.~\ref{forces_vs_eta}.} \label{energy_correction}
\end{figure}

Next, we look at the total energy. According to Eq.~\ref{tot_energy}, the total energy should depend on both the hardness and the charge response kernel. This is born out, as shown in Fig.~\ref{energy_correction}a. Therefore, one needs to resolve this discrepancy by rewriting the total energy expression in terms of $\Delta\boldsymbol{\nu}$ and $\boldsymbol{\chi}$ only.

As shown previously~\cite{scalfi2020charge},  the following equality holds under the variational condition:
\begin{equation}
    \mathbf{c}\T 
    \boldsymbol{\eta} \mathbf{c} = - \Delta \boldsymbol{\nu}\T \mathbf{c} \label{qAq_eq_bq}
\end{equation}

Thus we can replace $\mathbf{c}\T \boldsymbol\eta \mathbf{c}$ with $- \Delta \boldsymbol{\nu}\T \mathbf{c} $ and add a correction to the total energy as
\begin{equation}
    \Delta U = -\frac{1}{2}[\mathbf{c}\T \boldsymbol\eta \mathbf{c} + \Delta \boldsymbol\nu \T \mathbf{c} ] \label{correction_def}
\end{equation}

If the $\boldsymbol\eta-\boldsymbol\chi$ relation as defined by Eq.~\ref{chi_def} is fulfilled, this term should be 0. A non-zero term arises when they are not self-consistent.

When applying this correction, the total energy does not depend anymore on the hardness used by the MD engine, as expected (Fig.~\ref{energy_correction}b).  Thus, we now have everything checked to run MD properly with a ML derived CRK via the PiNNwall interface.

\subsection{Benchmarking on perfect metal electrode}

\begin{figure}[ht]
\begin{center}
\includegraphics[width=0.8\linewidth]{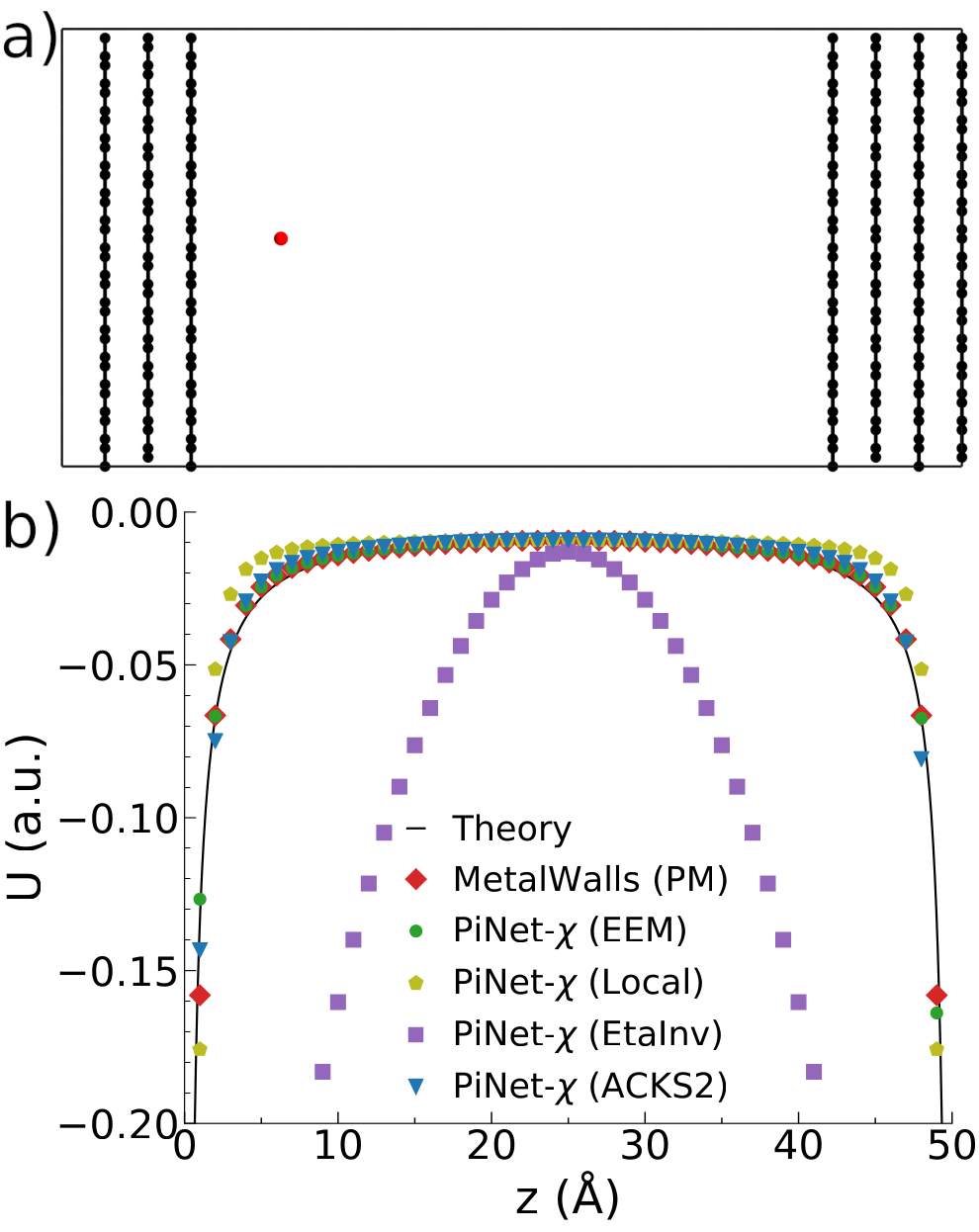}
\end{center}
\caption{\textbf{The electrostatic energy of a test charge between two sides of a graphite electrode}. a) The graphite electrode in vacuum under 3D PBCs is used as the model for representing a perfect metal electrode. b) The total electrostatic energy of the system when moving the test charge between two sides of electrode. The solid line corresponds to the theoretical result $U=-\dfrac{q\epsilon_0}{\mathrm{z}}-\dfrac{q\epsilon_0}{L-\mathrm{z}}$, where L is the size of the vacuum slab and z the distance between the test charge and the electrode surface. \label{test_capacitor}}
\end{figure}

As a first test, a unit charge is put on top of the middle of a carbon ring of the interfacial plane and moved in the vacuum space between the two planes (Fig.~\ref{test_capacitor}a). The total energy as a function of the charge position for the different models (MetalWalls, ACKS2, EEM, EtaInv and Local) is displayed on Fig.~\ref{test_capacitor}b along with the theoretical line. The ACKS2 and EEM are found more close to the theoretical line, which makes them the candidates for the next test. \added{Note that MetalWalls (PM) throughout this work refers to simulations done with the default Gaussian width parameters as implemented in the code and originated from the work of Reed, Lanning and Madden~\cite{reed_electrochemical_2007}}.

\begin{figure}[ht]
\begin{center}
\includegraphics[width=0.8\linewidth]{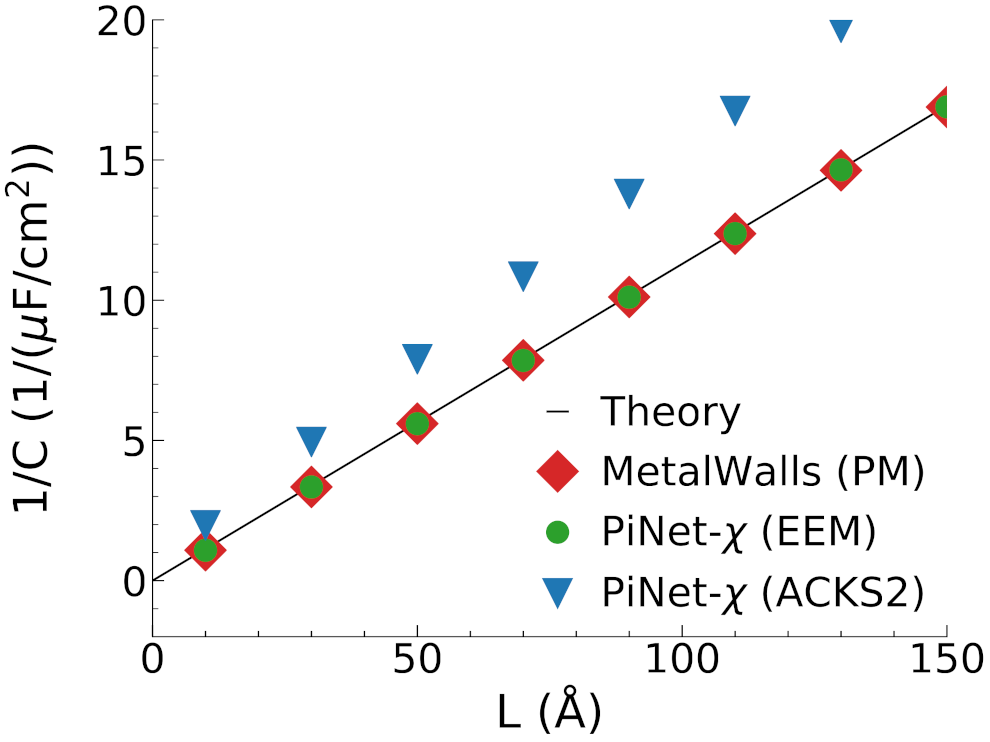}
\end{center}
\caption{\textbf{The inverse capacitance $1/C$ of an empty capacitor as a function of the vacuum slab size $L$}. The system under consideration is the one shown in Fig.~\ref{test_capacitor}a (without the test charge). The solid line corresponds to the theoretical result  $1/C=L/\epsilon_0$\label{test_empty_capacitor}}
\end{figure}
In the second test, we used the same graphite system as in Fig.~\ref{test_capacitor}a and computed the corresponding capacitance by varying the size of the vacuum slab. When the graphite model behaves like a PM with the dielectric constant of infinity, the total capacitance will be only determined by the size of the vacuum.  Its capacitance for the different models (MetalWalls, ACKS2, and EEM) as a function of the electrode separation is computed by applying a finite-field that leads to a potential bias of 2V, and the results are displayed on Fig.~\ref{test_empty_capacitor}. In this case, the EEM kernel shows a metallic behavior and follows almost exactly the theoretical line, as compared to ACKS2. The results of ACSK2 indicate that the electric field inside the graphite model is finite, which leads a smaller polarization and a lower integral capacitance. 

Based on these tests, we will employ the EEM kernel generated from PiNet-$\chi$ in the following case studies of chemically doped graphene and graphene oxide electrodes.
In order to separate the effects of the local geometry and the chemical heterogeneity on polarizability, we will also employ a PiNet-$\chi$ model by considering all the atoms as carbon atoms for the computation of the CRK, which is referred as PiNet-$\chi$ (EEM all C).

\section{Application to chemically doped graphene and graphene oxide electrodes}

\subsection{Nitrogen-doped graphene electrode}
\begin{figure}[ht]
\begin{center}
\includegraphics[width=0.8\linewidth]{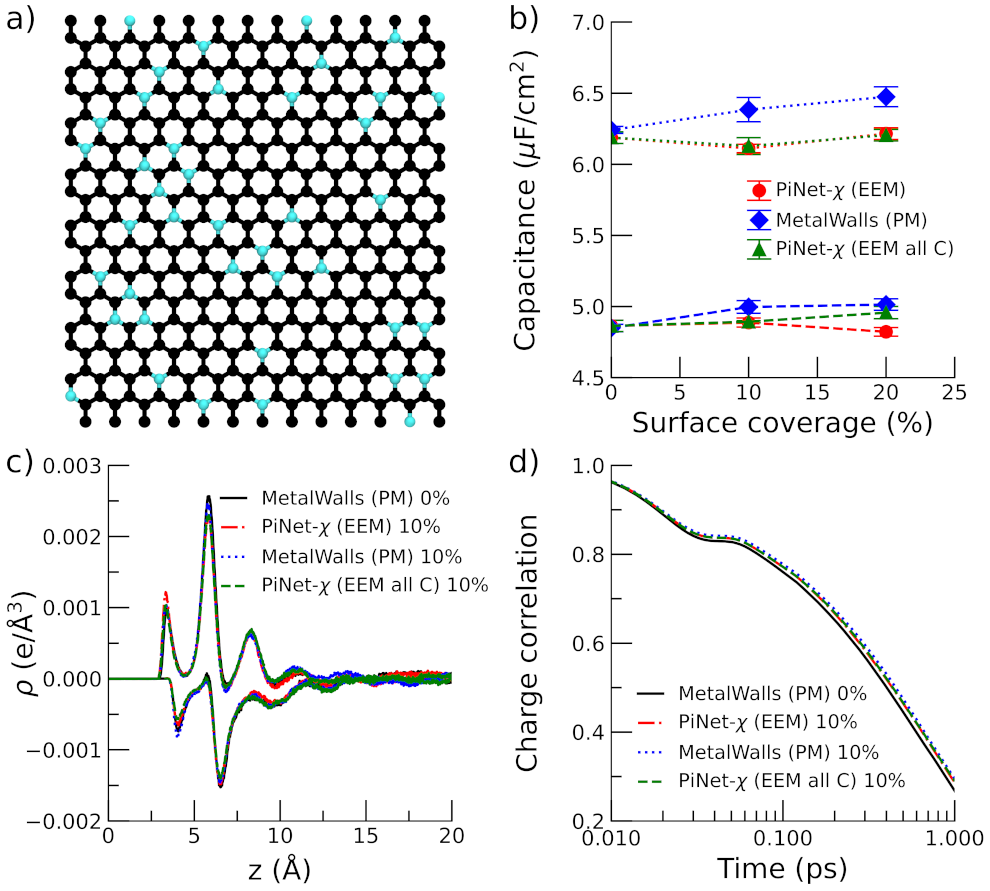}
\end{center}
\caption{\textbf{Nitrogen-doped graphene electrode}. a) Snapshot of the electrode surface with a 10\% surface coverage. (electrolyte solution is not shown for clarity). b) Helmholtz capacitance for the positive and negative electrodes as a function of the surface coverage. Dashed lines correspond to the positive electrode while dotted lines  correspond to the negative electrode. c) Total charge density of ionic species as a function of the distance to the negative/positive electrode under an applied potential of 2V. The distance is taken from the position of the carbon plane. d) Time correlation function of the electrode charge under an applied potential of 0V. \label{graphitic_results}}
\end{figure}

The simplest way to introduce chemical heterogeneity in the graphene layers is through the chemical doping, such as nitrogen, which shows a significant improvement on electrochemical activities~\cite{wang2010, wang2018}. Due to its valence, nitrogen substitution does not induce an out-of-plane change in the layer structure itself (Fig.~\ref{graphitic_results}a).

It is found that substituting carbon by nitrogen has a very limited impact on the Helmholtz capacitance (Fig.~\ref{graphitic_results}b). This is also reflected in the charge density profile of ions next to the electrode as well as the dynamics of electrode charge (Fig.~\ref{graphitic_results}c and Fig.~\ref{graphitic_results}d respectively).

We also notice that regardless of model, the asymmetry in the Helmholtz capacitance between the positive and negative electrode remains, in which the capacitance of the negative electrode has a much higher capacitance at the same surface density. This is in accord with the observation that the cation distribution is more close to the electrode surface than that of anions. 

\subsection{Graphene oxide electrode with epoxy terminations}

\begin{figure}[ht]
\begin{center}
\includegraphics[width=0.8\linewidth]{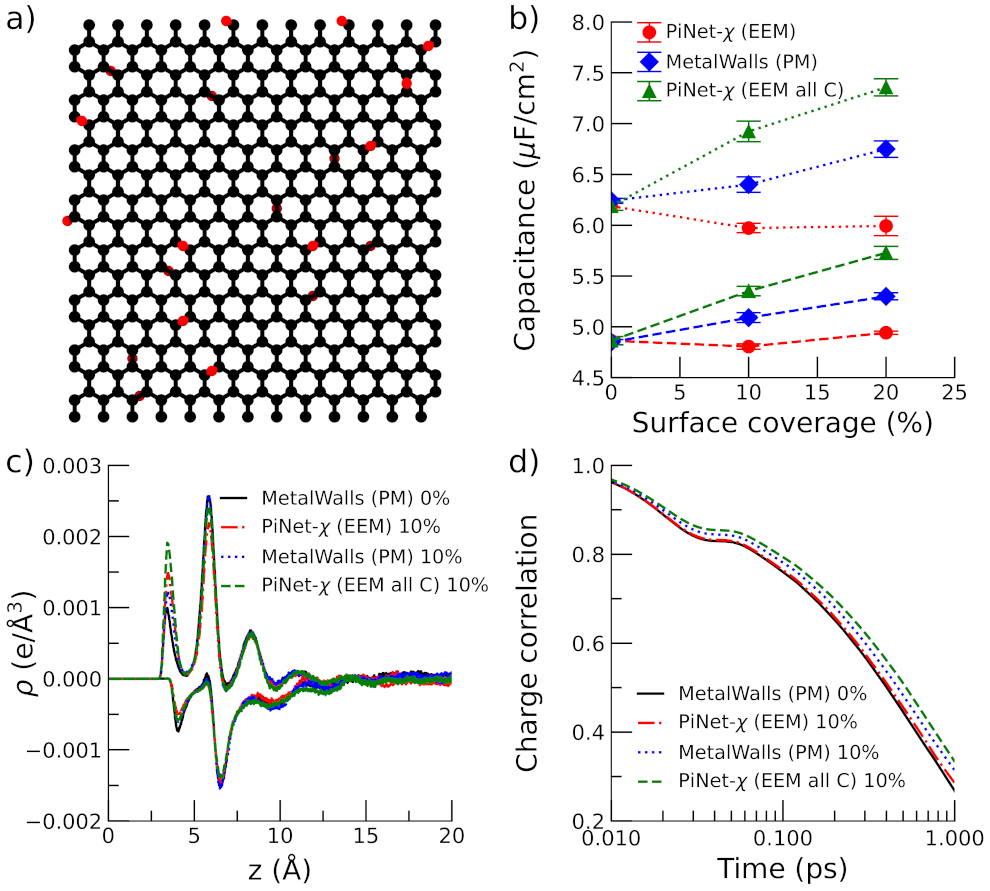}
\end{center}
\caption{\textbf{Graphene oxide electrode with epoxy terminations}. a) Snapshot of the electrode surface with a 10\% surface coverage. (electrolyte solution is not shown for clarity). b) Helmholtz capacitance for the positive and negative electrodes as a function of the surface coverage. Dashed lines correspond to the positive electrode while dotted lines  correspond to the negative electrode. c) Total charge density of ionic species as a function of the distance to the negative/positive electrode under an applied potential of 2V. The distance is taken from the position of the carbon plane. d) Time correlation function of the electrode charge under an applied potential of 0V. \label{epoxy_results}}
\end{figure}

Epoxy, hydroxyl, and carboxylic acid functional groups are commonly found in the graphene oxide~\cite{chen12}. In this section, we will look at how the Helmholtz capacitance will change upon introducing epoxy termination in the graphene oxide. This adds one layer of complexity as it also changes the roughness of the surface (Fig.~\ref{epoxy_results}a).

In contrast to the case of the graphitic substitution as shown in the previous section, the doping with oxygen under the form of epoxy groups will modify the capacitance significantly (Fig.~\ref{epoxy_results}b). Both PiNet-$\chi$ (EEM all C) and MetalWalls (PM) treat electrode atoms as carbon atoms regardless of element types, and yet PiNet-$\chi$ (EEM all C) shows a more rapid increment in the capacitance with the surface coverage as compared to MetalWalls (PM). This highlights the fact that the CRK implemented in PiNet-$\chi$ does take into account the change in the ``metallicity'' due to the local geometry. 

When comparing PiNet-$\chi$ (EEM all C) and PiNet-$\chi$ (EEM), the effect of chemical heterogeneity in the polarizability at atomic site comes into play.  This in turn decreases the capacitance, due to a smaller polarizability of oxygen and hydrogen atoms as compared to that of carbon atoms. Therefore, the gain in the capacitance due to the surface roughness and the local geometry is cancelled out by introducing the chemical heterogeneity. 

As shown in Fig.~\ref{epoxy_results}c and Fig.~\ref{epoxy_results}d, the charge density profiles of ions  and the correlation function of the electrode charge do correlate with the observed capacitance. For instance, PiNet-$\chi$ (EEM all C), which has the highest capacitance, shows a strongest first peak of charge density for both positive and negative electrodes and the longest relaxation time. Nevertheless, this correlation is not perfect, in which the first peak height of charge density next to the negative electrode does no decrease in the same order as that in its capacitance. This suggests that the ion population in the second peak of charge density also contributes to the resulting capacitance.  

\subsection{Graphene oxide electrode with hydroxyl terminations}

\begin{figure}[ht]
\begin{center}
\includegraphics[width=0.8\linewidth]{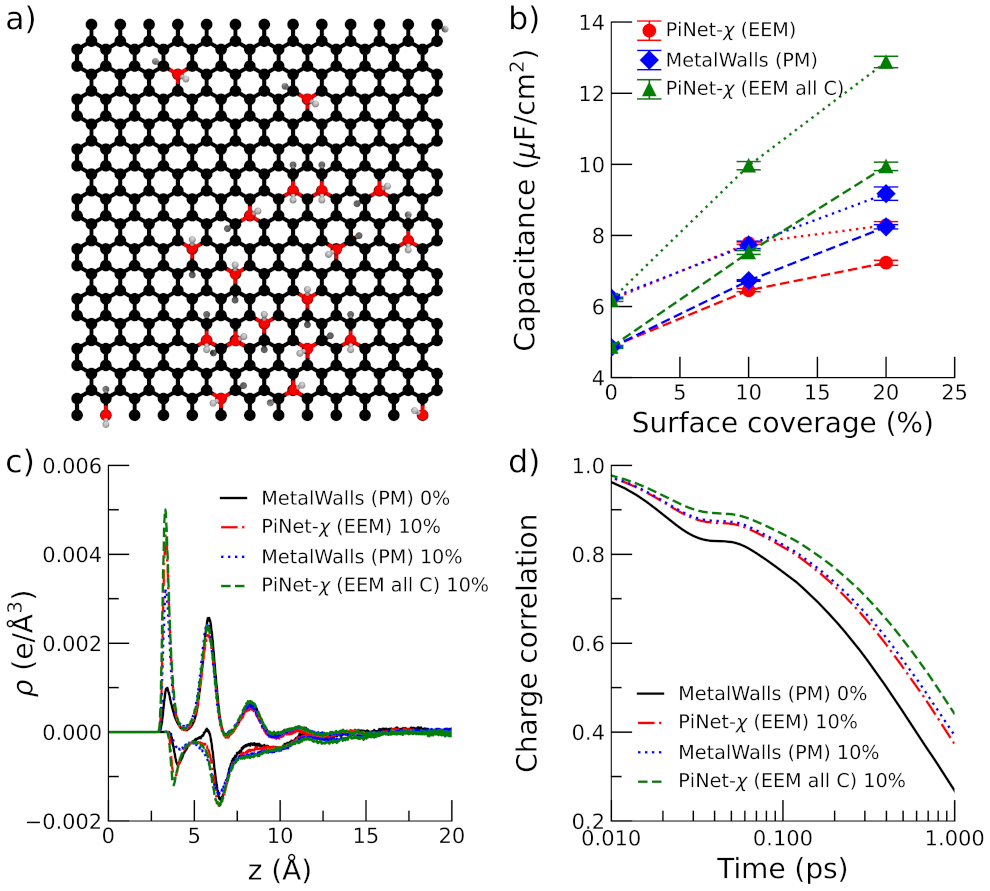}
\end{center}
\caption{\textbf{Graphene oxide electrode with hydroxyl terminations}. a) Snapshot of the electrode surface with a 10\% surface coverage. (electrolyte solution is not shown for clarity). b) Helmholtz capacitance for the positive and negative electrodes as a function of the surface coverage. Dashed lines correspond to the positive electrode, while dotted lines correspond to the negative electrode. c) Total charge density of ionic species as a function of the distance to the negative/positive electrode under an applied potential of 2V. The distance is taken from the position of the carbon plane. d) Time correlation function of the electrode charge under an applied potential of 0V.  \label{hydroxyl_results}}
\end{figure}

Next, we also looked into the case of the hydroxyl terminated graphene oxide, as shown in Fig.~\ref{hydroxyl_results}a.

In general, the trends for the capacitance (Fig.~\ref{hydroxyl_results}b), the charge density profile of ions (Fig.~\ref{hydroxyl_results}c) as well as the time correlation function of the electrode charge (Fig.~\ref{hydroxyl_results}d) look similar to those observed in the case of the epoxy terminated graphene oxide. Nevertheless, there are also considerable differences between the two cases. The capacitance obtained in the case of the hydroxyl terminated graphene oxide is much higher than the epoxy case for the same surface coverage. Notably, the corresponding charge densities of ions at both positive and negative electrodes also have much higher intensities (Fig.~\ref{hydroxyl_results}c). This suggests that by increasing the surface coverage of OH groups, the electrode surface becomes more hydrophilic and ion populations next to electrode surface increase because of a more favorable solvation environment. 

\subsection{Graphene oxide with proton charge}

Examples in previous sections focus on the interplay between the geometrical effect on metallicity and the chemical heterogeneity in polarizabilty by comparing the perfect metal model in MetalWalls, PiNet-$\chi$ (EEM) and PiNet-$\chi$ (EEM all C). In this section, we will apply PiNet-$\chi$ (EEM) to probe the surface acid-base chemistry of electrode materials instead.

\begin{figure}[ht]
\begin{center}
  \includegraphics[width=0.8\linewidth]{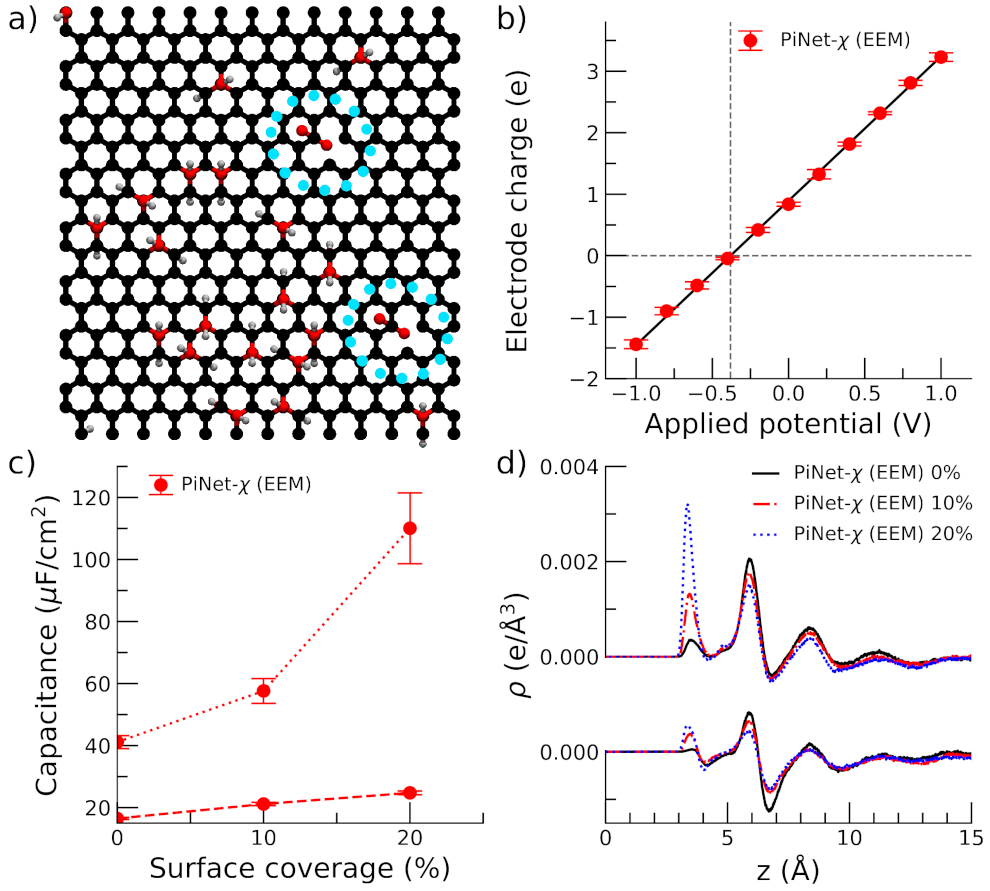}
\end{center}
\caption{\textbf{Graphene oxide with proton charge}. a) Snapshot of the carboxyl-terminated electrode surface with a 10\% surface coverage of OH. (electrolyte solution is not shown for clarity and the location of deprotonated carboxyl groups are highlighted.). b) The electrode charge as a function of the applied potential. $V_\textrm{PZFC}$ is identified when the electrode charge becomes zero. c) Helmholtz capacitance for the positive and negative electrodes as a function of the surface coverage of OH. Dashed line corresponds to the positive electrode, while dotted line corresponds to the negative electrode. d) Total charge density of ionic species as a function of the distance to the negative/positive electrode at the point of zero free charge (PZFC). The distance is taken from the position of the carbon plane.   \label{carboxyl_results}}
\end{figure}

In graphene oxide,  both surface carboxylic and hydroxyl groups can undergo protonation/deprotonation depending on the solution pH. It has been reported that the $p$K$_a$ is about 6.6 for the carboxylic group and 9.8 for the hydroxyl group in graphene oxide~\cite{10.1021/jz300236w}. This means that, at the neutral pH, the most relevant ionizable group in graphene oxide is the carboxylic group and the most probable acid-base reaction is the one shown in Eq.~\ref{eq:pKa2}. Therefore, in this section, we will explore the PiNNwall interface for modelling protonic double layer at the graphene oxide surface terminated with carboxylic groups (Fig.~\ref{carboxyl_results}a) 

\begin{eqnarray}
\label{eq:pKa2}
    \textrm{GO}-\mathrm{COOH} \mathrm{(aq)} &\rightleftharpoons & \textrm{GO}-\mathrm{COO}^{-} \mathrm{(aq)} + \mathrm{H}^{+} \mathrm{(aq)} \\
\label{eq:pKa1}
\textrm{GO}-\mathrm{COOH}_{2}^{+} \mathrm{(aq)} &\rightleftharpoons &
    \textrm{GO}-\mathrm{COOH}  \mathrm{(aq)} +  \mathrm{H}^{+} \mathrm{(aq)}
\end{eqnarray}

As shown in Fig.~\ref{carboxyl_results}b, by changing the applied potential, one can identify the point of zero free charge (PZFC) due to the electronic polarization. This ``titration'' procedure is similar to the one used before in modelling charged insulator/electrolyte interfaces for eliminating the finite-size effect~\cite{Zhang:2016ca}. It is worth noting that the slope of Fig.~\ref{carboxyl_results}b yields a capacitance of value 4.7 $\mu F/\textrm{cm}^2$, which is comparable to that of pristine graphene (see Fig.~\ref{graphitic_results}b for the case of 0\% surface coverage). 

Once the PZFC is identified, the integral capacitance can be computed readily using the $dq/dV_\textrm{PZFC}$ formula, in which $q$ is the proton charge that we introduced through the protonation and deprotonation of carboxyl groups. The result of the computed Helmholtz capacitance due to the proton charge at the PZFC is shown in Fig.~\ref{carboxyl_results}c. What is surprising is that the resulting Helmholtz capacitance for the hydroxylated surface with deprotonated carboxyl groups can be as large as 100 $\mu F/\textrm{cm}^2$. This is one order magnitude higher compared to those found in pristine graphene but very similar in magnitude as those reported for metal oxide~\cite{zhang2019coupling,jia2021origin}. Therefore, this finding provides a clue why the Helmholtz capacitance found in metal oxide is much higher than that found in the metal, as often seen in experiments~\cite{Lyklema1991-ch}.

\section{Conclusion and outlook}

In this work, we have integrated the atomistic ML code (PiNN) and the MD simulation code (MetalWalls) to model heterogeneous electrode surfaces. PiNN was used to generate the response kernel and the base charge from ML models PiNet-$\chi$ and PiNet-dipole respectively. Then, this information was passed to the MetalWalls to carry out efficient computations of electrostatic interactions and to propagate the dynamics. 

Through validation and verification, we have identified PiNet-$\chi$ (EEM) as the candidate for practical applications, which shows almost identical results for pure carbon electrode compared to the original Siepmann-Sprik model. Thanks to the flexibility of PiNet-$\chi$ (EEM) for modelling any electrode materials composed of C, N, O, H, S, Cl, we were able to study both chemically doped graphene electrode and graphene oxide with various terminations. 

It is found that while the surface roughness and hydrophilicity can potentially increase the capacitance, these beneficial effects are attenuated by a smaller polarizability of elements (N, O, and H) involved in the chemical heterogeneity. On the other hand, we showed that the proton charge due to the surface acid-base chemistry at graphene oxide surfaces can lead to a significant increment in capacitance, which is comparable in magnitude (100 $\mu F/\textrm{cm}^2$) to those reported in metal oxide-based systems.

Given that the capacitance is so different depending on whether the electronic or the protonic charge dominates, it would be interesting to study the transition between these two cases in future works, which can shed light on the electrochemical behavior of the ``polarized oxide surfaces''~\cite{Ardizzone:1996ca}. Reparameterizing PiNet-$\chi$ for transition metal oxides or transition metal dichalcogenides would allow us to investigate an even broader range of complex electrode materials in contact with both aqueous and non-aqueous electrolytes. In terms of the development of PiNNwall, future works can also be considered in the direction to pass the forces from PiNN to MetalWalls. In combination with ML potential for modelling the electrode materials~\cite{Deringer:2017ea,2020.Rowe, Lombardo22}, this will enable us to study the electrode dynamics and its role in the electrochemical energy storage.

\section*{Acknowledgements}
This project has received funding from the European Research Council (ERC) under the European Union's Horizon 2020 research and innovation programme (grant agreement No. 949012). L.K. is partly supported by a PhD studentship from the Centre for Interdisciplinary Mathematics (CIM) at Uppsala University. The
  simulations were performed on the resources provided by the National Academic
  Infrastructure for Supercomputing in Sweden (NAISS) at PDC partially funded by the Swedish Research Council through grant agreement No. 2022-06725 and through the project access to the LUMI supercomputer, owned by the EuroHPC Joint Undertaking, hosted by CSC (Finland) and the LUMI consortium.

\section*{Supporting Information}
  Further validation of passing the charge response kernel, details of the validation and the implementation of base charges predicted from PiNet-dipole, and ion distributions at the electrified interfaces.

 \bibliographystyle{achemso} 
 %\bibliography{report}

\providecommand{\latin}[1]{#1}
\makeatletter
\providecommand{\doi}
  {\begingroup\let\do\@makeother\dospecials
  \catcode`\{=1 \catcode`\}=2 \doi@aux}
\providecommand{\doi@aux}[1]{\endgroup\texttt{#1}}
\makeatother
\providecommand*\mcitethebibliography{\thebibliography}
\csname @ifundefined\endcsname{endmcitethebibliography}
  {\let\endmcitethebibliography\endthebibliography}{}

\end{document}

%\twocolumn
\maketitle
%thispagestyle{empty}

\makeatletter 
\renewcommand{\thefigure}{S\@arabic\c@figure} \renewcommand{\thetable}{S\@arabic\c@table}
\renewcommand{\theequation}{S\@arabic\c@equation} 
\renewcommand{\thepage}{S\arabic{page}}

\makeatother

\appendix{} 
\setcounter{figure}{0} 
\setcounter{table}{0}

\centerline{\bf\Large{Supporting Information}}

\section{Further validation of passing the charge response kernel}
In the Figure 2 of the Main Text, we observed a difference in the response charge predictions of PiNN and MetalWalls using the same CRK, that we attributed to the use of point charge-point charge electrostatic interaction with the test charge for the former and Gaussian charge-point charge interaction for the later. Here we show a validation of this hypothesis by using the same setup as in the Figure 2 of the Main Text but with a Gaussian width of 0.2~\AA ~instead.  Results are shown on Figure S1. Indeed, one can see that the difference in response charges predicted by PiNN and Metawalls become negligible for all cases by decreasing the Gaussian width. 

\begin{figure}[H]
\begin{center}
\includegraphics[width=0.9\linewidth]{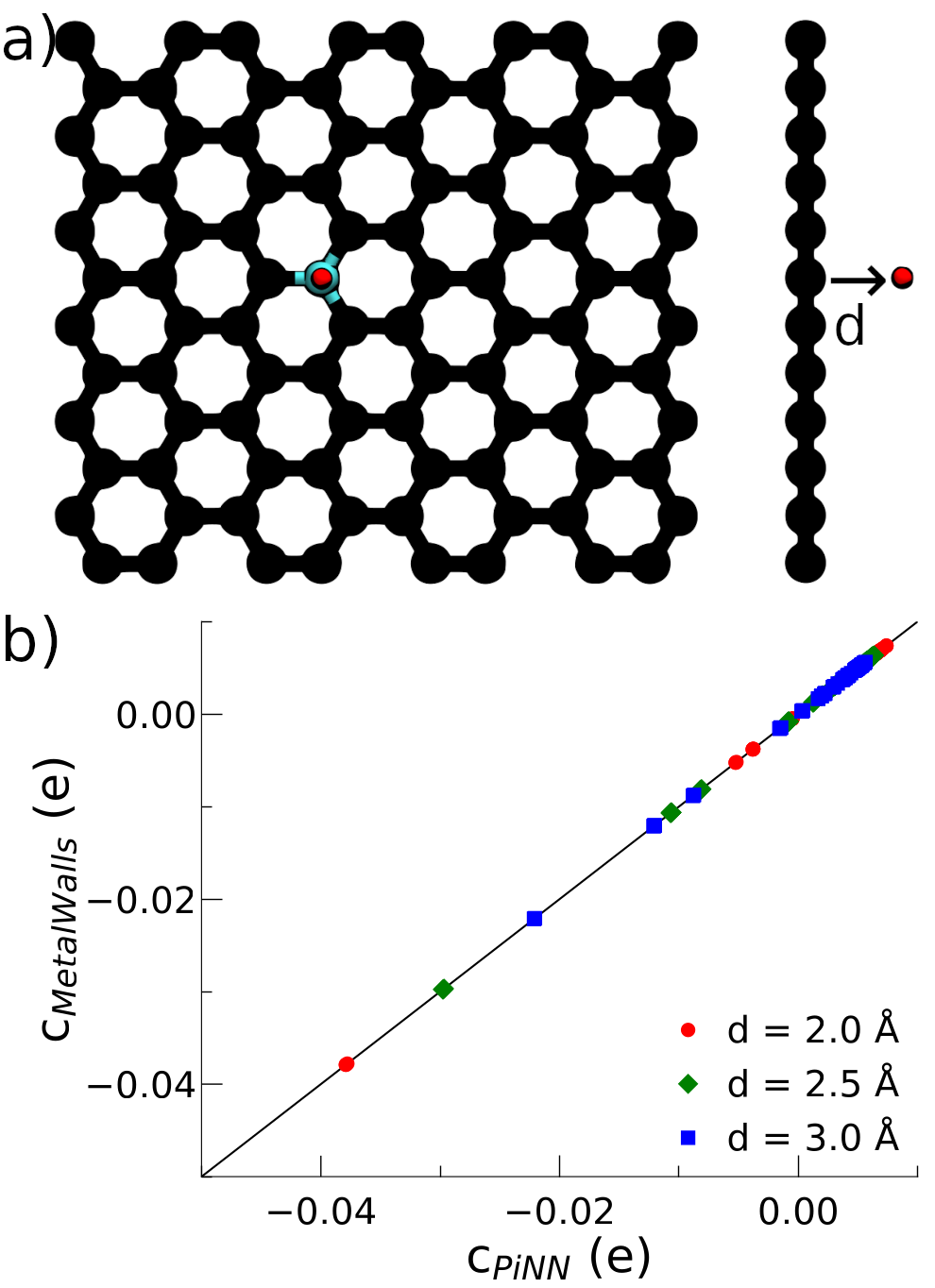}
\end{center}
\caption{\textbf{Passing the charge response kernel}. Response charges predicted by MetalWalls via the PiNNwall interface against the prediction from PiNN using the same kernel PiNet-$\chi$ (EEM).}
\end{figure}

\newpage 

\newpage 

\section{Validation and implementation of base charges predicted from PiNet-dipole}
To validate the charges predicted from the PiNet-dipole model, molecular analogues of the target structures were used for each of the functionalized graphene models. To validate that the charges predicted from PiNet-dipole have a physical basis, comparisons were made to charges computed using several population analysis techniques.

To perform the population analysis, DFT calculations were run using Gaussian09 ~\cite{gaussian}. The B3LYP~\cite{becke1992density,stephens1994ab} functional and the cc-pVDZ basis set \cite{dunning1989gaussian} were used. The population analyses that were performed are: CM5 \cite{marenich2012charge}, Mulliken~\cite{mulliken1955electronic}, Hirshfeld~\cite{hirshfeld1977bonded} and Merz-Singh-Kollman (MSK)~\cite{Singh84, besler1990atomic}. The molecular analogues were deemed fit as a references if the predicted charges corresponded to chemical intuition and the dipole moment was comparable to that calculated using DFT. 

For the graphene sheet doped with nitrogen, a planar form of trimethylamine was used as a reference.

\begin{figure}[H]
\begin{center}
\includegraphics[width=\linewidth]{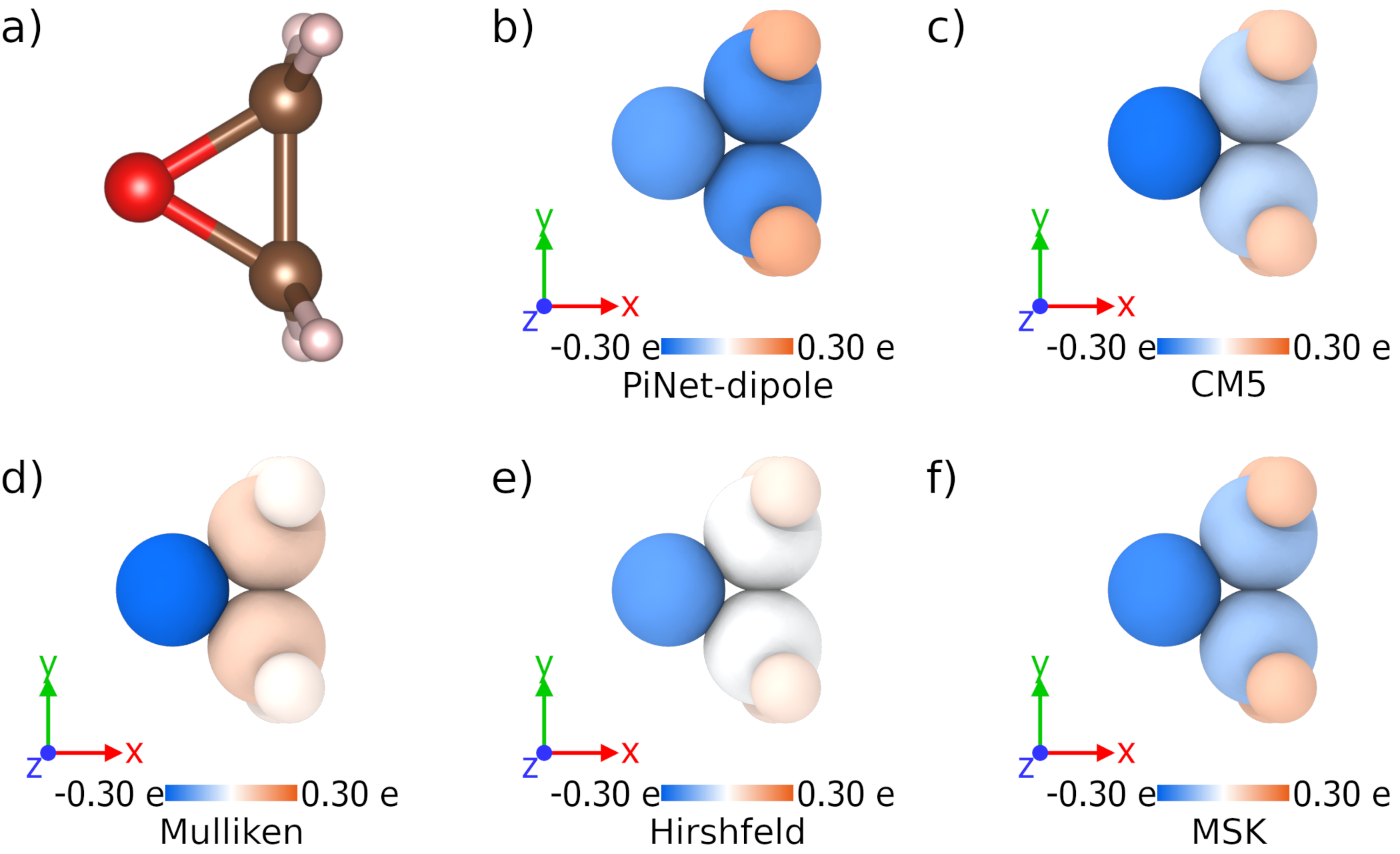}
\end{center}
\caption{\textbf{Charges of trimethylamine}. The structure of trimethylamine (a). Computed with b) PiNet-dipole, c) CM5, d) Mulliken, e) Hirshfeld, and f) MSK.   \label{trimethylamine_charges}}
\end{figure}

\begin{table}[H]
\begin{center}
\begin{tabular}{lrrr}
\hline
Method & D$_{x}$ (D) & D$_{y}$ (D) & D$_{z}$ (D)\\
\hline 
DFT & -0.340 & -0.286 & -0.010 \\
PiNet-dipole & -0.435 & -0.169 & -0.016 \\
CM5 & -0.126 & -0.123 & -0.013 \\
MSK & -0.320 & -0.280 & -0.009 \\
\hline
\end{tabular}
\caption{\textbf{Dipole moment of trimethylamine}.}
\end{center}
\end{table}

\begin{table}[H]
\begin{center}
\begin{tabular}{lr}
\hline
Element & PiNet charge (e)\\
\hline 
N & -0.19527 \\
C & 0.03033 \\
C & 0.08597 \\
C & 0.07899 \\
\hline
\end{tabular}
\caption{\textbf{Base charges from trimethylamine as implemented in the N-doped graphene}.}
\end{center}
\end{table}

The charges of the methyl groups are placed the carbon atoms when charges are used in the real system, to ensure a charge neutral entity. 

For the graphene sheet doped with epoxy groups, ethylene oxide was used as the reference molecule.

\begin{figure}[H]
\begin{center}
\includegraphics[width=\linewidth]{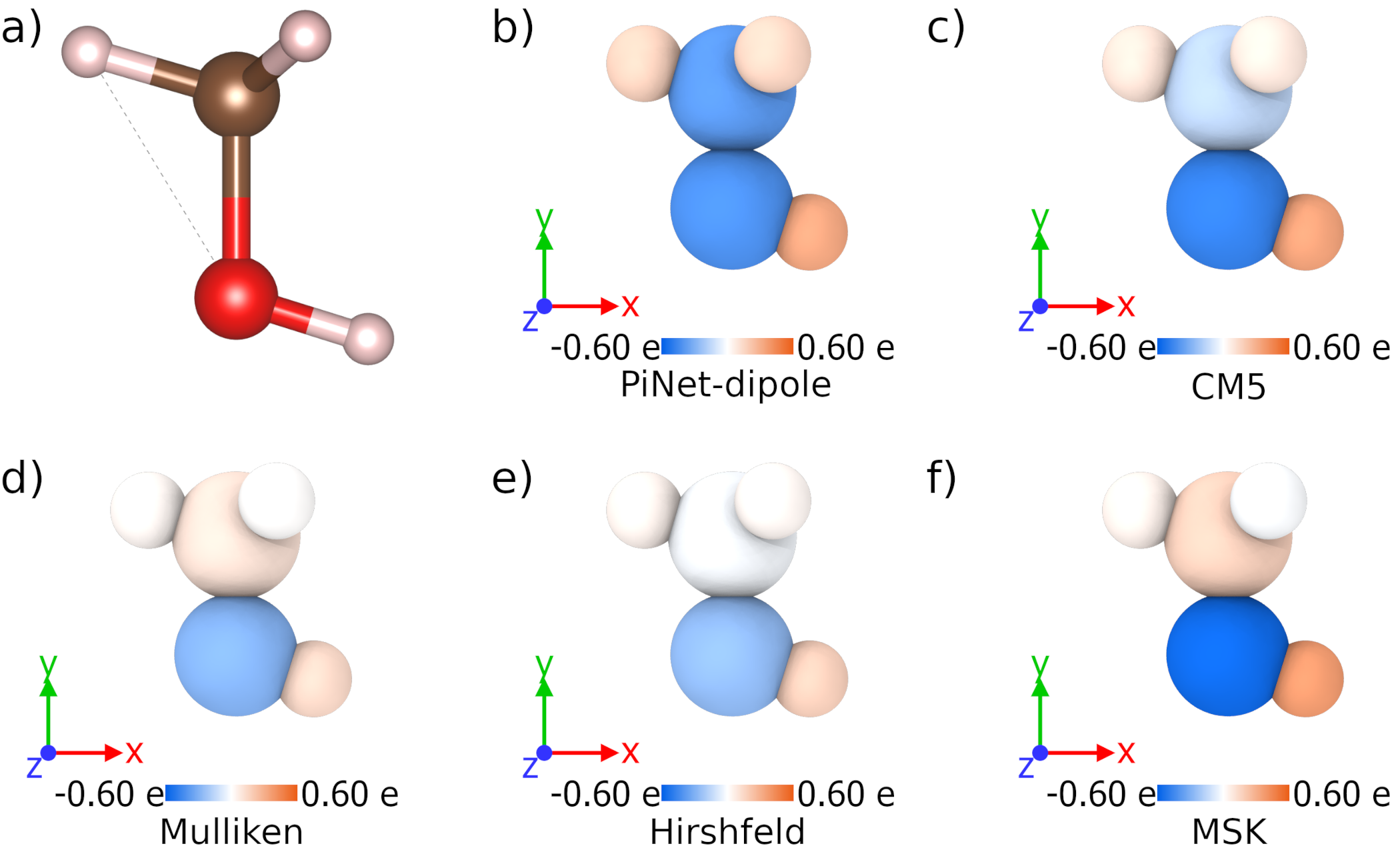}
\end{center}
\caption{\textbf{Charges of ethylene oxide}. The structure of ethylene oxide (a). Computed with b) PiNet-dipole, c) CM5, d) Mulliken, e) Hirshfeld, and f) MSK.   \label{ethylene_oxide_charges}}
\end{figure}

\begin{table}[H]
\begin{center}
\begin{tabular}{lrrr}
\hline
Method & D$_{x}$ (D) & D$_{y}$ (D) & D$_{z}$ (D)\\
\hline 
DFT & 1.806 & 0.000 & -0.141 \\
PiNet-dipole & 1.811 & 0.000 & -0.142 \\
CM5 & 2.052 & 0.000 & -0.161 \\
MSK & 1.837 & 0.000 & -0.144 \\
\hline
\end{tabular}
\caption{\textbf{Dipole moment of ethylene oxide}.}
\end{center}
\end{table}

\begin{table}[H]
\begin{center}
\begin{tabular}{lr}
\hline
Element & PiNet charge (e)\\
\hline 
O & -0.19297 \\
C & 0.09648 \\
C & 0.09648 \\
\hline
\end{tabular}
\caption{\textbf{Based charges from ethylene oxide as implemented in the epoxy-terminated graphene oxide}.}
\end{center}
\end{table}

Here, the charges of the hydrogen atoms are also combined with that of the carbon atoms when the charges are transferred to functionalized graphene. Once again, to ensure charge neutrality and to localize the charges on the graphene sheet. 

For the graphene sheet doped with hydroxyl groups, methanol was used as the molecular analogue.

\begin{figure}[H]
\begin{center}
\includegraphics[width=\linewidth]{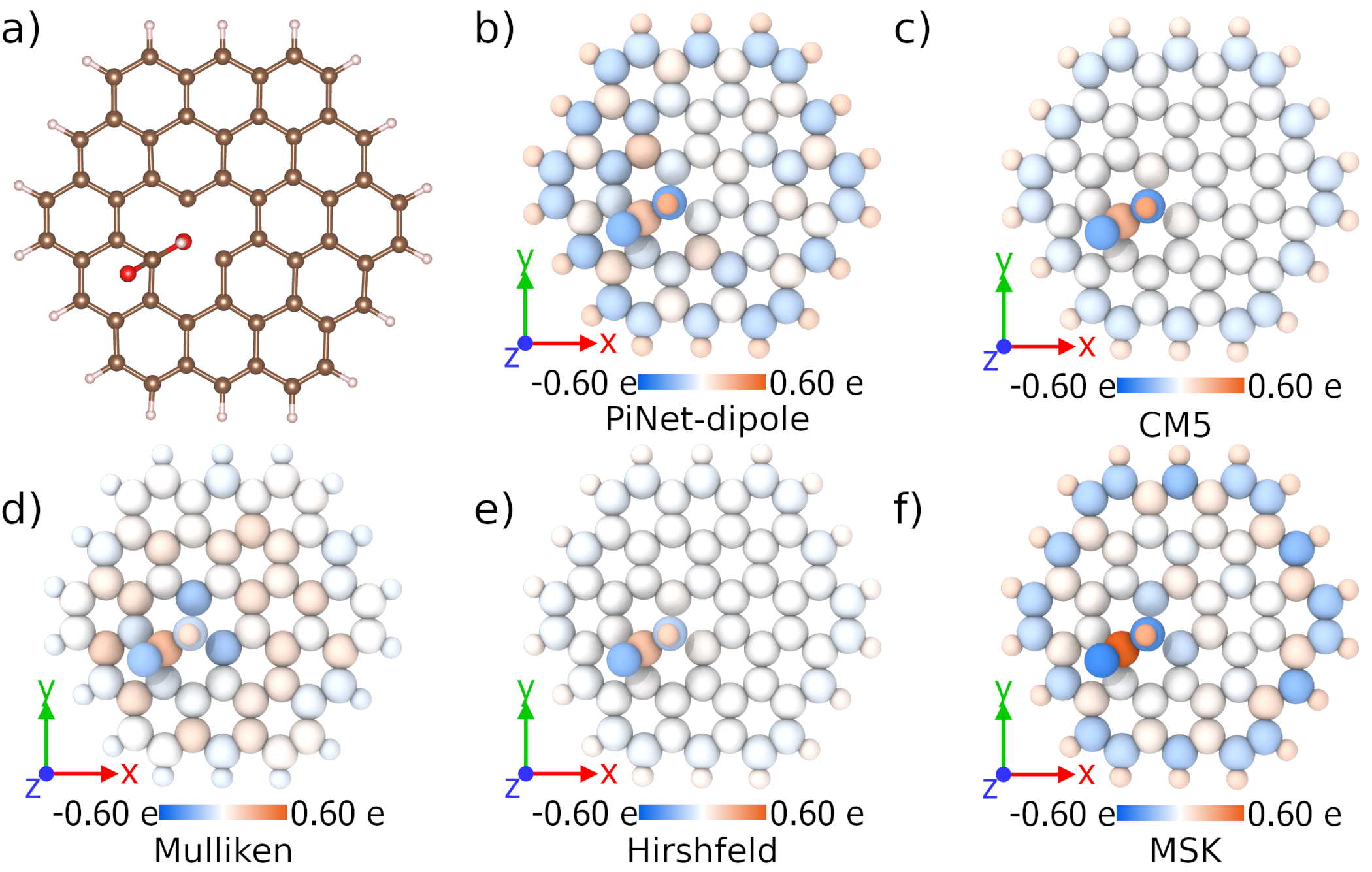}
\end{center}
\caption{\textbf{Charges of methanol}. The structure of methanol (a). Computed with b) PiNet-dipole, c) CM5, d) Mulliken, e) Hirshfeld, and f) MSK.   \label{methanol_charges}}
\end{figure}

\begin{table}[H]
\begin{center}
\begin{tabular}{lrrr}
\hline
Method & D$_{x}$ (D) & D$_{y}$ (D) & D$_{z}$ (D)\\
\hline 
DFT & 1.340 & 0.825 & 0.000 \\
PiNet-dipole & 1.327 & 1.067 & 0.000 \\
CM5 & 1.384 & 0.957 & 0.000 \\
MSK & 1.326 & 0.851 & 0.000 \\
\hline
\end{tabular}
\caption{\textbf{Dipole moment of methanol}.}
\end{center}
\end{table}

\begin{table}[H]
\begin{center}
\begin{tabular}{lr}
\hline
Element & PiNet charge (e)\\
\hline 
O & -0.41668 \\
H & 0.32197 \\
C & 0.09471 \\
\hline
\end{tabular}
\caption{\textbf{Base charges from methanol as implemented in the hydroxyl-terminated graphene oxide}.}
\end{center}
\end{table}

The charge on the carbon atom is set so that it includes the charges of the hydrogen atoms as well. In this way, it compensates for the charge on the hydroxyl group, and the whole group is charge neutral.

Finally, for the graphene sheet functionalized with carboxyl groups, a smaller graphene flake with carboxyl groups was used as a reference. This was done because alternative analogues showed large fluctuations in the charges when changing the charge state of the analogue which did not correspond to chemical intuition. While the dipole moment for the carboxyl flake shows discrepancies to that of DFT in the $x$- and $y$-direction, the $z$-direction, which is the most important direction when it comes to the carboxyl group, agrees within a reasonable error margin. 

\begin{figure}[H]
\begin{center}
\includegraphics[width=\linewidth]{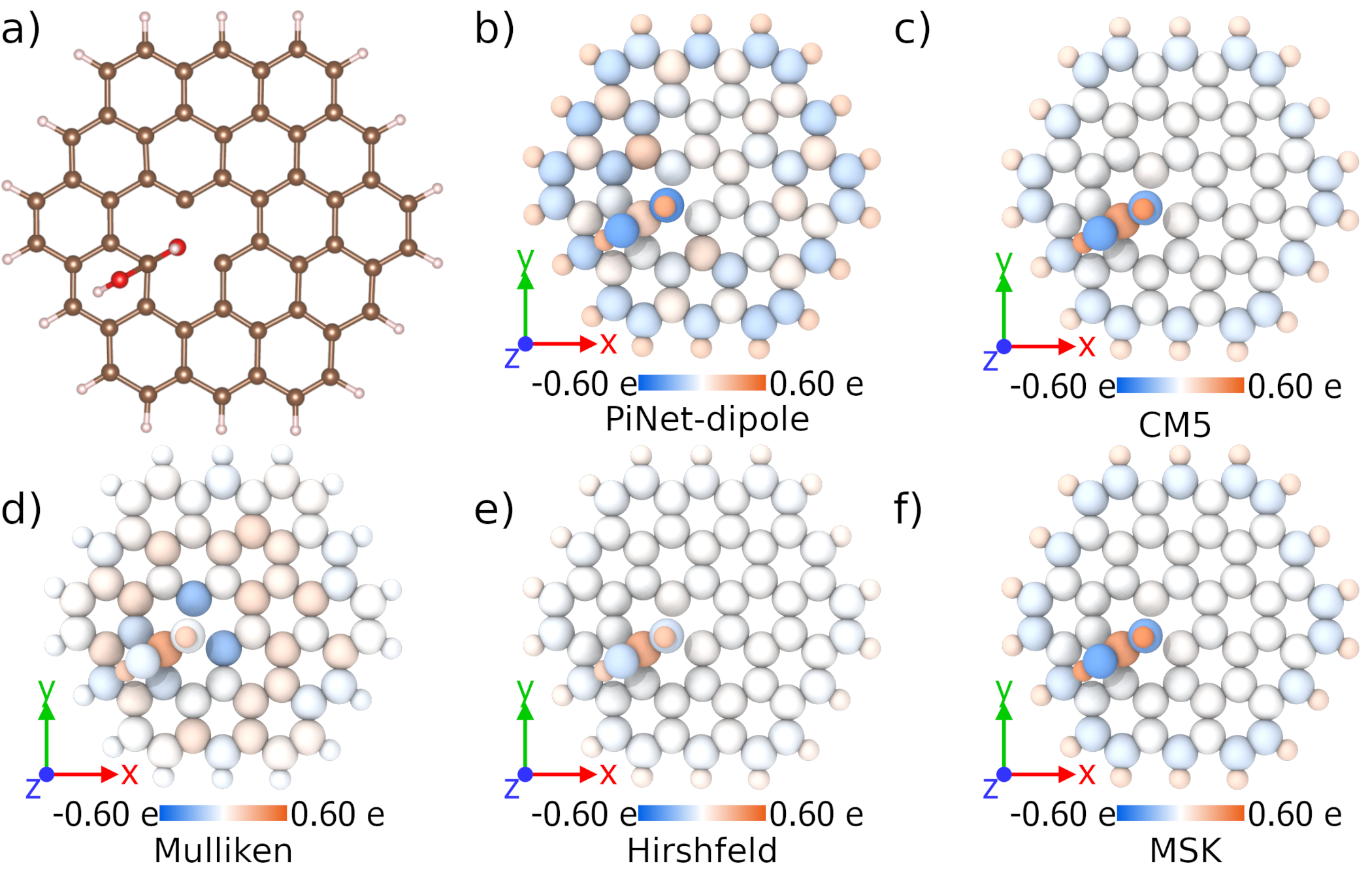}
\end{center}
\caption{\textbf{Charges of the neutral carboxyl flake}. The structure of the neutral carboxyl flake (a). Computed with b) PiNet-dipole, c) CM5, d) Mulliken, e) Hirshfeld, and f) MSK.   \label{carboxyl_neutral_charges}}
\end{figure}

\begin{table}[H]
\begin{center}
\begin{tabular}{lr}
\hline
Element & PiNet charge (e)\\
\hline 
O$_{\mathrm{double-bonded \ O}}$ & -0.28245 \\
O$_{\mathrm{OH}}$ & -0.31622 \\
H$_{\mathrm{OH}}$ & 0.32579 \\
C & 0.27288 \\
\hline
\end{tabular}
\caption{\textbf{Charges from the neutral carboxyl flake}.}
\end{center}
\end{table}

\begin{table}[H]
\begin{center}
\begin{tabular}{lrrr}
\hline
Method & D$_{x}$ (D) & D$_{y}$ (D) & D$_{z}$ (D)\\
\hline 
DFT & 0.2953 & 0.0873 & 0.3845 \\
PiNet-dipole & 3.301 & 1.699 & 0.145 \\
CM5 & 0.435 & 0.230 & 0.297 \\
MSK & 0.265 & 0.087 & 0.354 \\
\hline
\end{tabular}
\caption{\textbf{Dipole moment from the neutral carboxyl flake}.}
\end{center}
\end{table}

\begin{figure}[H]
\begin{center}
\includegraphics[width=\linewidth]{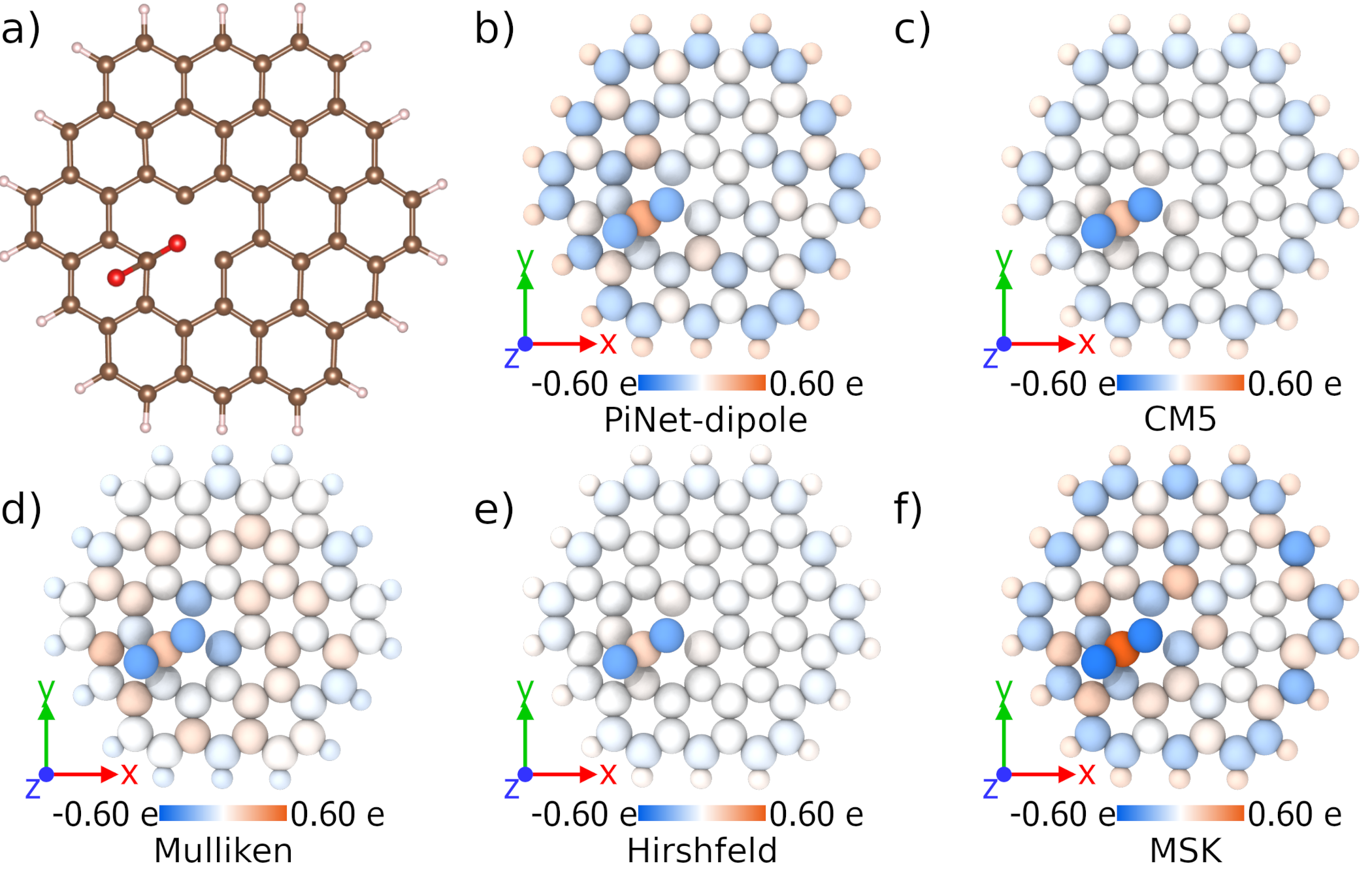}
\end{center}
\caption{\textbf{Charges of the  protonated carboxyl flake}. The structure of the protonated carboxyl flake (a). Computed with b) PiNet-dipole, c) CM5, d) Mulliken, e) Hirshfeld, and f) MSK.   \label{carboxyl_protonated_charges}}
\end{figure}

\begin{table}[H]
\begin{center}
\begin{tabular}{lr}
\hline
Element & PiNet charge (e)\\
\hline 
O$_{\mathrm{double-bonded \ O}}$ & -0.33339 \\
H$_{\mathrm{double-bonded \ O}}$ & 0.27653 \\
O$_{\mathrm{OH}}$ & -0.36729 \\
H$_{\mathrm{OH}}$ & 0.33747 \\
C & 1.08668 \\
\hline
\end{tabular}
\caption{\textbf{Base charges from the protonated carboxyl flake as implemented in the protonated side of carboxyl-terminated graphene oxide}.}
\end{center}
\end{table}

Since the investigated structures contain neutral, protonated, and deprotonated forms of the carboxyl groups, these are also the structures for which the charges were predicted. Here the charge of the carbon atom is set such that the total charge of the protonated carboxyl group is +1. Once again the charges of the atoms are predicted using PiNet-dipole. Then, the charge of the carbon atom is simply set to ensure that the charge of the protonated carboxyl group sums to +1. This is done to keep the charges localized, and because it is the simplest way to adjust the charge without the need for an arbitrary charge division. It also prevents unphysical modifications to the other charges from being made. This is supported by Figure \ref{carboxyl_protonated_charges}, as this shows that the charge analysis performed with DFT methods the excess charge is also mostly located on the carbon atom.

\begin{figure}[H]
\begin{center}
\includegraphics[width=\linewidth]{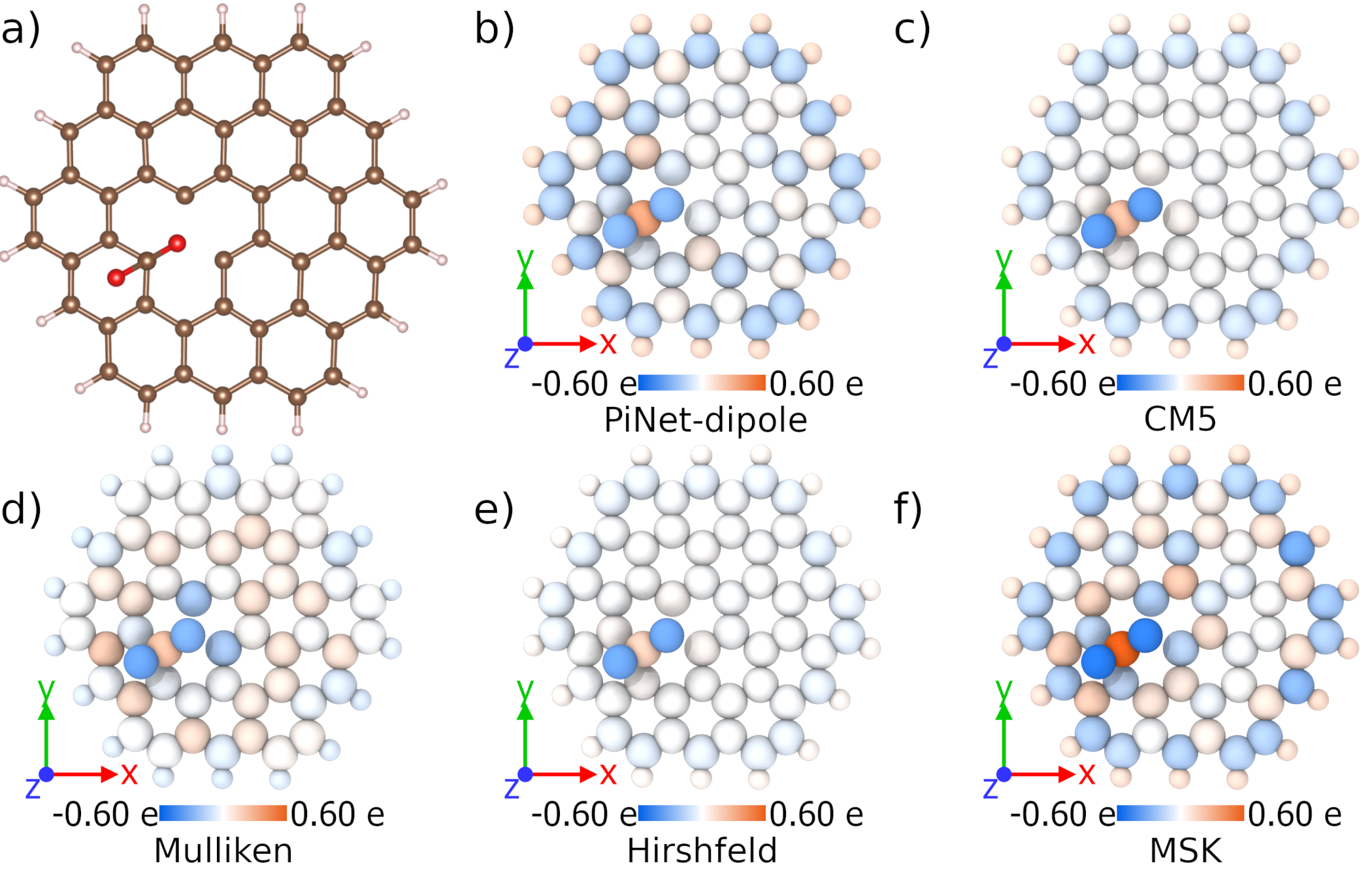}
\end{center}
\caption{\textbf{Charges of the deprotonated carboxyl flake}. The structure of the deprotonated carboxyl flake (a). Computed with b) PiNet-dipole, c) CM5, d) Mulliken, e) Hirshfeld, and f) MSK.   \label{carboxyl_deprotonated_charges}}
\end{figure}

\begin{table}[H]
\begin{center}
\begin{tabular}{lr}
\hline
Element & PiNet charge (e)\\
\hline 
O$_{\mathrm{double-bonded \ O}}$ & -0.28599 \\
O$_{\mathrm{OH}}$ & -0.28723 \\
C & -0.42678 \\
\hline
\end{tabular}
\caption{\textbf{Base charges from the deprotonated carboxyl flake as implemented in the deprotonated side of carboxyl-terminated graphene oxide}.}
\end{center}
\end{table}

Similarly, for the deprotonated cases, the charge of the carbon atom is set such that the total charge of the deprotonated carboxyl group is -1. As can be seen in Figure \ref{carboxyl_deprotonated_charges}, for the DFT charge methods the negative charge is spread across the carboyxl flake, mostly at the edges. As a first approximation, the excess charge is localized on the carbon atom in our implementation, which avoids any size-inconsistent charge divisions.  

\newpage

\section{Ion distributions at the electrified interfaces}

\begin{figure}[ht]
\begin{center}
\includegraphics[width=0.8\linewidth]{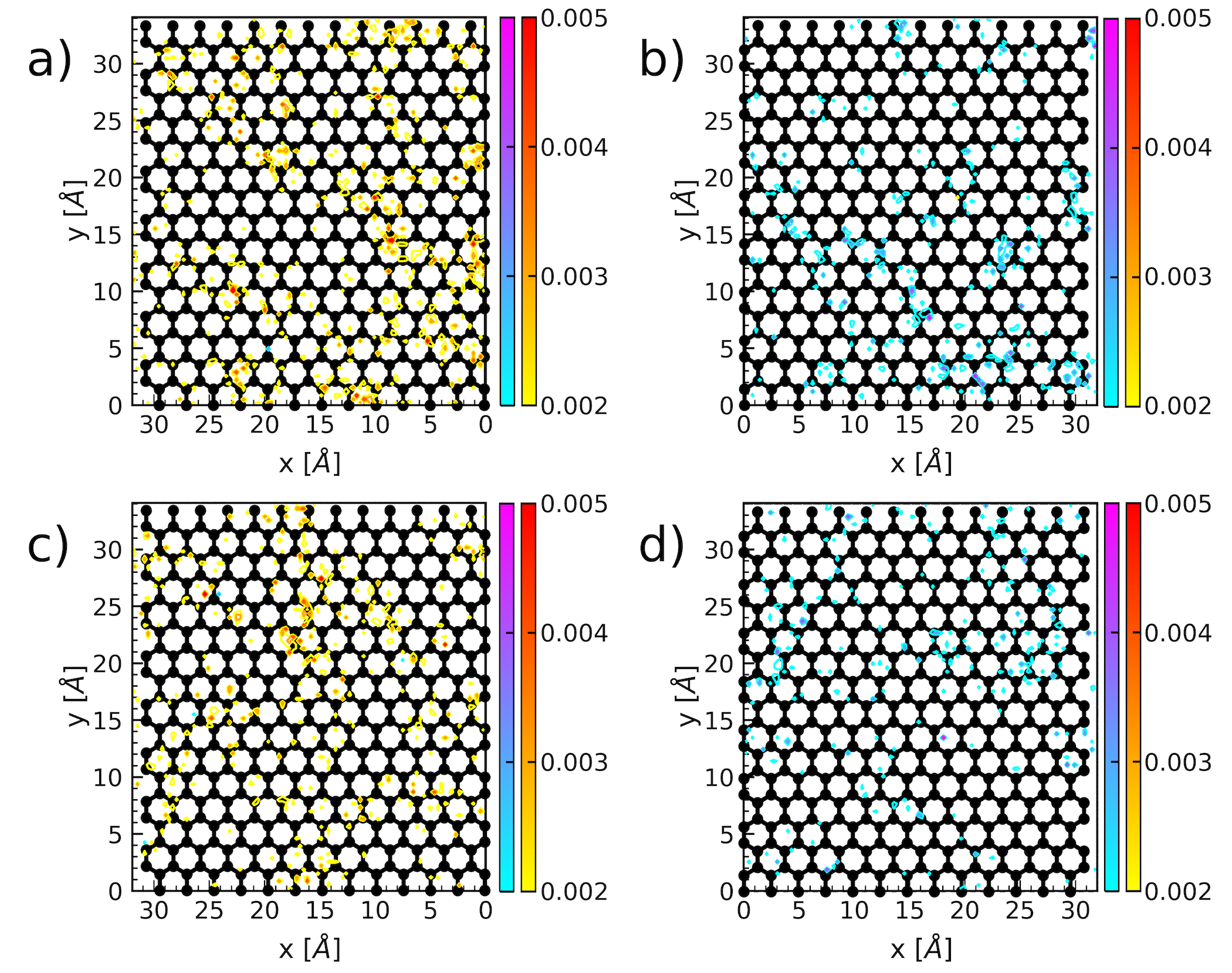}
\end{center}
\caption{\textbf{Density of adsorbed ions in on a pristine graphene electrode }. Surface density in e\AA$^{-2}$ of potassium (yellow to red color bar) and chloride (blue to purple color bar) using the MetalWalls (PM) on the negative (a) and positive (b) electrode, and using the PiNet-$\chi$ (EEM) model on the negative (c) and positive (d) electrode, under an applied potential of 2 V. \label{pristine_ions}}
\end{figure}

\begin{figure}[ht]
\begin{center}
\includegraphics[width=0.8\linewidth]{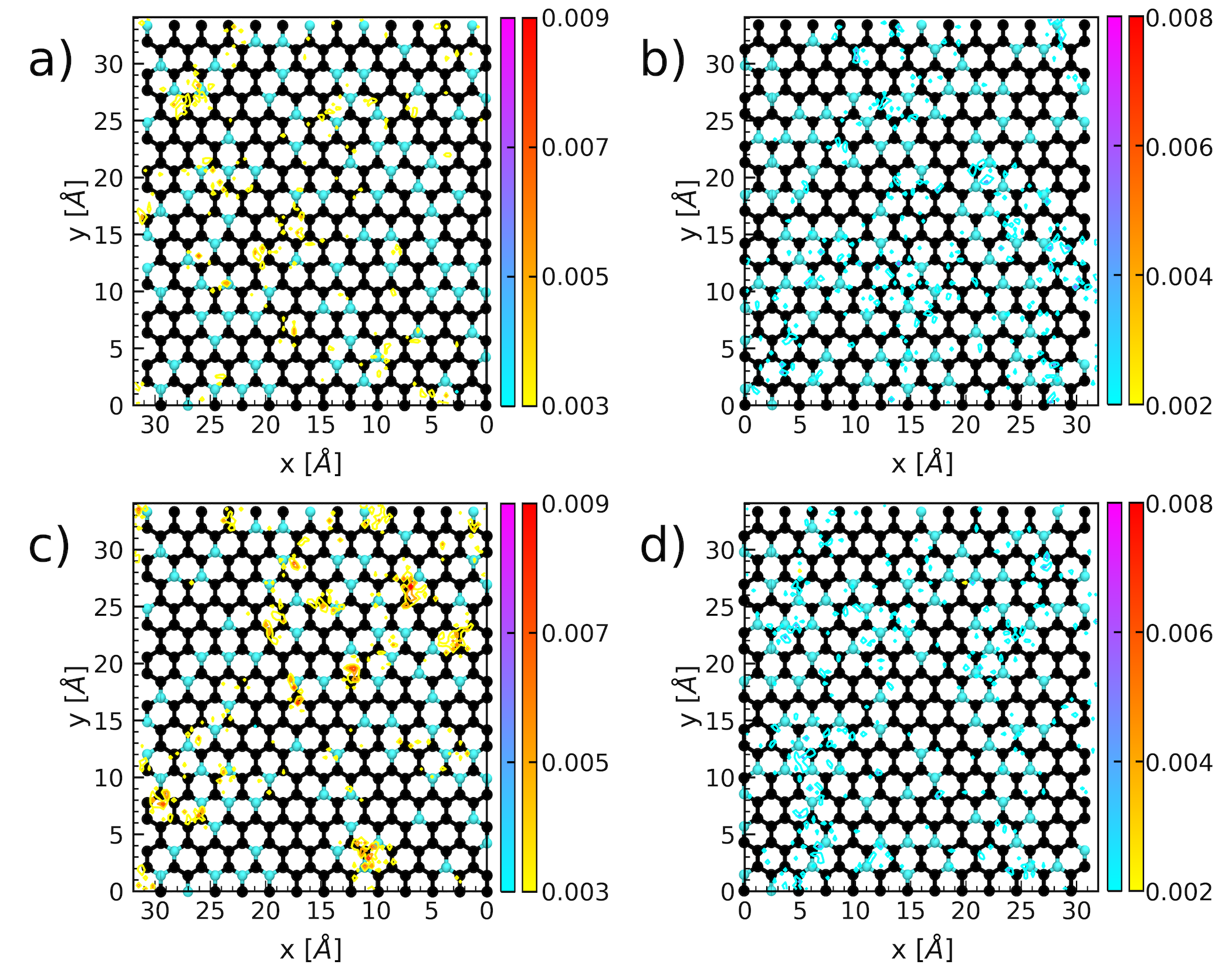}
\end{center}
\caption{\textbf{Density of adsorbed ions on a graphene electrode with Nitrogen substitution}. Surface density in e\AA$^{-2}$ of potassium (yellow to red color bar) and chloride (blue to purple color bar) using the MetalWalls (PM) model on the negative (a) and positive (b) electrode, and using the PiNet-$\chi$ (EEM) model on the negative (c) and positive (d) electrode, for a surface coverage of 20 \% and under an applied potential of 2 V. \label{graphitic_ions}}
\end{figure}

\begin{figure}[ht]
\begin{center}
\includegraphics[width=0.8\linewidth]{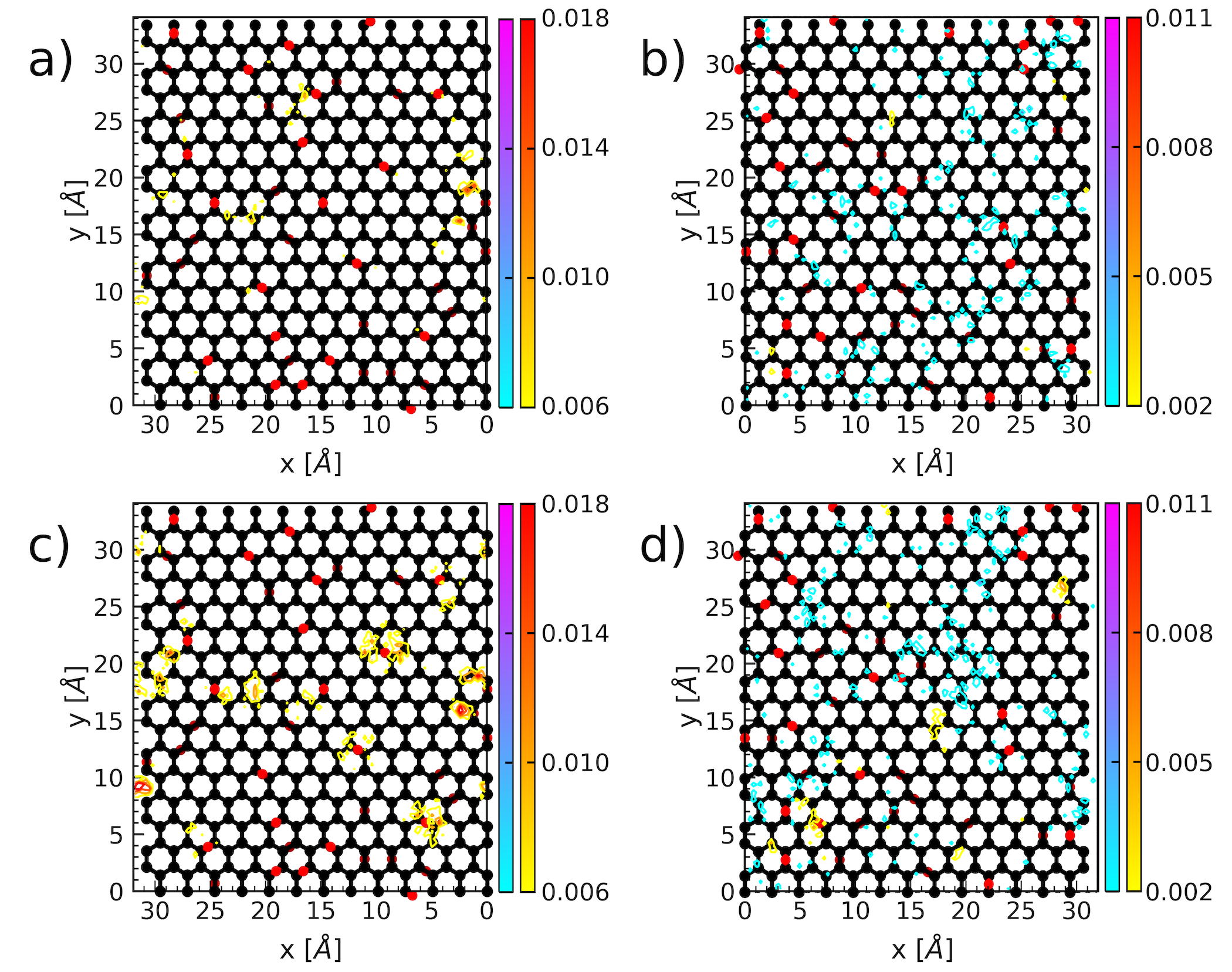}
\end{center}
\caption{\textbf{Density of adsorbed ions on a graphene oxide electrode with epoxy terminations}. Surface density in e\AA$^{-2}$ of potassium (yellow to red color bar) and chloride (blue to purple color bar) using the MetalWalls (PM) model on the negative (a) and positive (b) electrode, and using the PiNet-$\chi$ (EEM) model on the negative (c) and positive (d) electrode, for a surface coverage of 20 \% and under an applied potential of 2 V. \label{epoxy_ions}}
\end{figure}

\begin{figure}[ht]
\begin{center}
\includegraphics[width=0.8\linewidth]{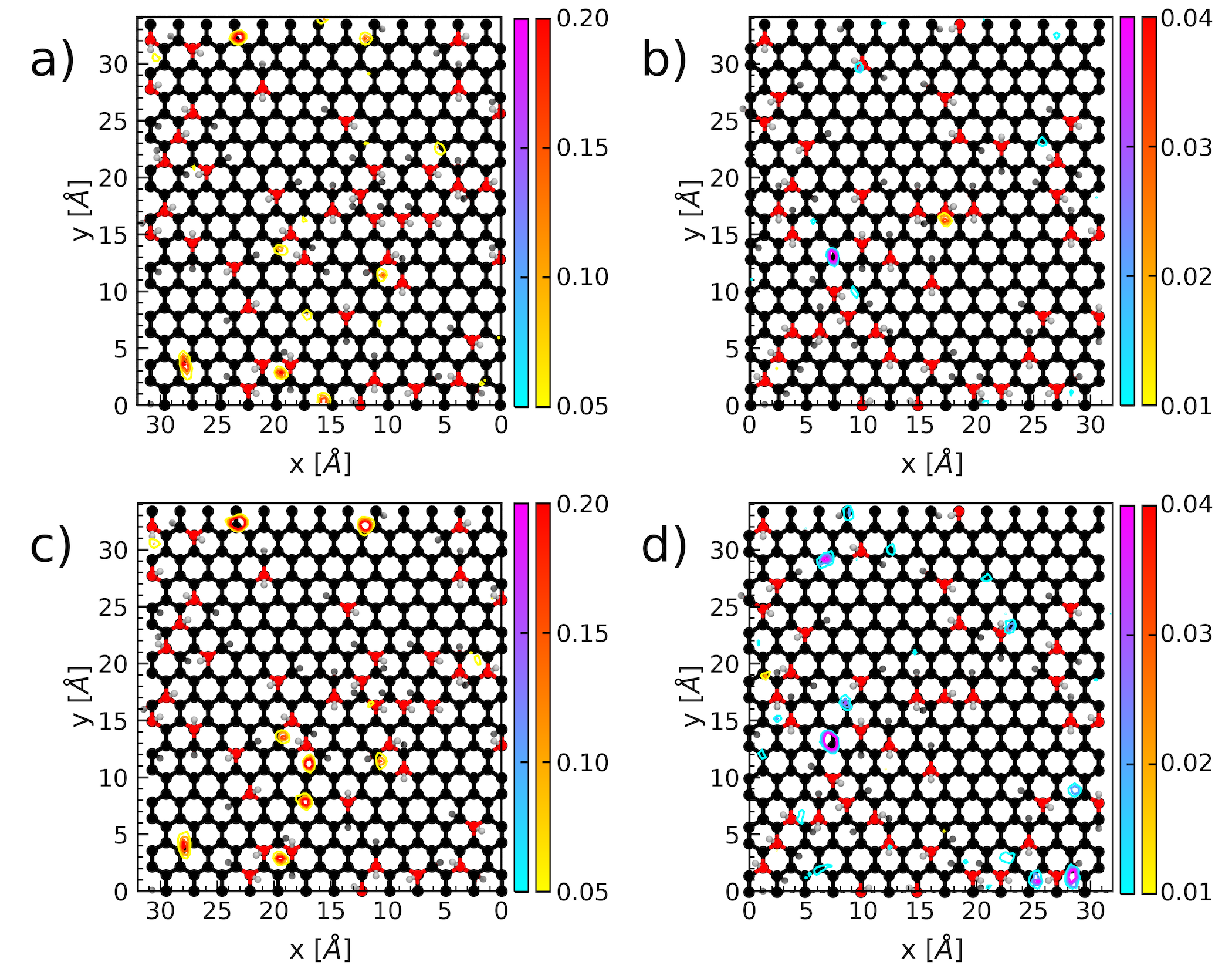}
\end{center}
\caption{\textbf{Density of adsorbed ions on a graphene oxide electrode with hydroxyl terminations}. Surface density in e\AA$^{-2}$ of potassium (yellow to red color bar) and chloride (blue to purple color bar) using the MetalWalls (PM) model on the negative (a) and positive (b) electrode, and using the PiNet-$\chi$ (EEM) model on the negative (c) and positive (d) electrode, for a surface coverage of 20 \% and under an applied potential of 2 V. \label{hydroxyl_ions}}
\end{figure}

\newpage

\bibliographystyle{acs}
%\bibliography{references.bib}
%\begin{thebibliography}{1}